\documentclass[10pt,aps,showpacs,nofootinbib,prd,aps,floats,
               amsmath,amssymb,amsfonts,floatfix,twocolumn]{revtex4-1}
\usepackage{amsmath, amssymb}
\usepackage{multirow}
\usepackage{paralist}
\usepackage{slashed}
\usepackage{hyperref} 
\newcommand{\mathsym}[1]{{}}

\usepackage{graphicx}
\usepackage{amsmath}
\usepackage{amssymb}
\usepackage{mathrsfs}
\newsavebox{\PSLASH}
 \sbox{\PSLASH}{$p$\hspace{-1.8mm}/}
 
\renewcommand{\theequation}{\thesection.\arabic{equation}}
\newcounter{saveeqn}
\newcommand{\add}{\addtocounter{equation}{1}}
\newcommand{\alphaeqn}{\setcounter{saveeqn}{\value{equation}}%
\setcounter{equation}{0}%
\renewcommand{\theequation}{\mbox{\thesection.\arabic{saveeqn}{\alpha{equation}}}}}
\newcommand{\reseteqn}{\setcounter{equation}{\value{saveeqn}}%
\renewcommand{\theequation}{\thesection.\arabic{equation}}}

 \newsavebox{\notrightarrow}
 \sbox{\notrightarrow}{$\to$\hspace{-4mm}/}
 
 \newsavebox{\PARTIALSLASH}
 \sbox{\PARTIALSLASH}{$\partial$\hspace{-1.6mm}/}
 
 \newsavebox{\ASLASH}
 \sbox{\ASLASH}{$A$\hspace{-2.1mm}/}
 
 \newsavebox{\KSLASH}
 \sbox{\KSLASH}{$k$\hspace{-1.8mm}/}
 
 \newsavebox{\LSLASH}
 \sbox{\LSLASH}{$\ell$\hspace{-1.8mm}/}
 
 \newsavebox{\QSLASH}
 \sbox{\QSLASH}{$q$\hspace{-1.8mm}/}
 
 \newsavebox{\DSLASH}
 \sbox{\DSLASH}{$D$\hspace{-2.2mm}/}
 
 \newsavebox{\DbfSLASH}
 \sbox{\DbfSLASH}{${\mathbf D}$\hspace{-2.8mm}/}
 
 \newsavebox{\DELVECRIGHT}
 \sbox{\DELVECRIGHT}{$\stackrel{\rightarrow}{\partial}$}
 
 \newcommand{\blue}{\IfColor{\textCadetBlue}{}}
\newcommand{\black}{\IfColor{\textBlack}{}}
\newcommand{\red}{\IfColor{\textRed}{}}
\newcommand{\green}{\IfColor{\textOliveGreen}{}}
\newcommand{\lil}{\IfColor{\textRedViolet}{}}

\newcommand{\bs}{\boldsymbol}

\usepackage{orcidlink}

\begin{document}
\title{Thermodynamic properties of a relativistic Bose gas under rigid rotation}
\author{E. Siri\,~\orcidlink{0009-0008-1021-3348}~~}\email{e.siri@physics.sharif.ir}\author{N. Sadooghi\,~\orcidlink{0000-0001-5031-9675}~~}\email{Corresponding author: sadooghi@physics.sharif.ir}
\affiliation{Department of Physics, Sharif University of Technology,
P.O. Box 11155-9161, Tehran, Iran}
\begin{abstract}
We study the thermodynamic properties of a rigidly rotating relativistic Bose gas. First, we derive the solution of the equation of motion corresponding to a rotating complex Klein-Gordon field and determine the free propagator of this model utilizing the Fock-Schwinger proper-time method.
Using this propagator, we then obtain the thermodynamic potential of this model in the zeroth and first perturbative level. In addition, we compute the nonperturbative ring contribution to this potential. Our focus is on the dependence of these expressions on the angular velocity, which effectively acts as a chemical potential. Using this thermodynamic potential, we calculate several quantities, including the pressure, angular momentum and entropy densities, heat capacity, speed of sound, and moment of inertia of this rigidly rotating Bose gas as functions of temperature ($T$), angular velocity ($\Omega$), and the coupling constant ($\alpha$). We show that certain thermodynamic instabilities appear at high temperatures and large couplings. They are manifested as zero and negative values of the above quantities, particularly the moment of inertia and heat capacity. Zero moment of inertia leads to the phenomenon of supervorticity at certain $T$ or $\alpha$. Supervortical temperatures (couplings) decrease with increasing coupling (temperature). We also observe superluminal sound velocities at high $T$ and for large $\alpha$.

\end{abstract}
\maketitle
\section{Introduction}\label{sec1}
\setcounter{equation}{0}
\begin{figure*}[hbt]
\includegraphics[width=11cm, height=6cm]{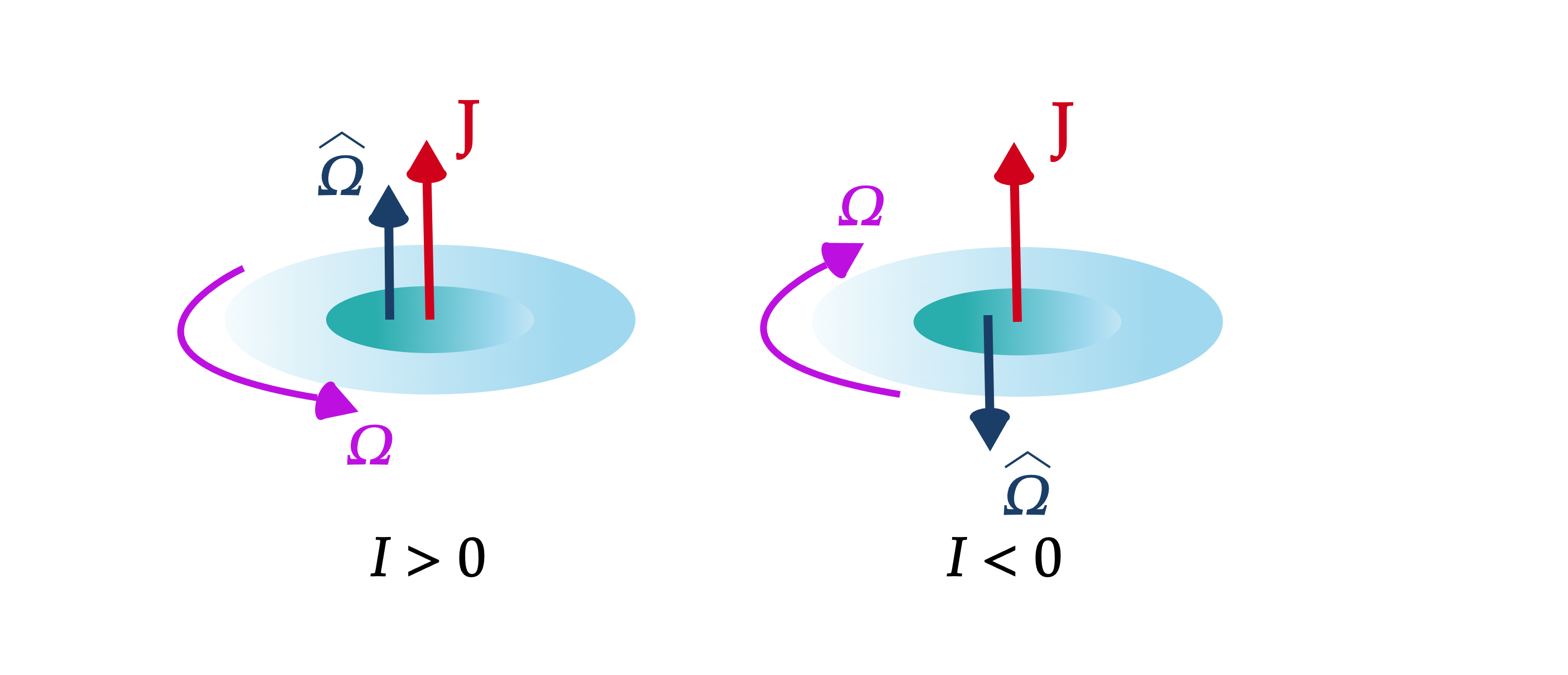}
\vspace{-1cm}
\caption{By applying an angular momentum $\boldsymbol{J}$, a system with a positive (negative) moment of inertia $I$ rotates with an angular velocity $\boldsymbol{\Omega}$ parallel (antiparallel) to $\boldsymbol{J}$.}\label{fig1}
\end{figure*}
Studying the effects of extreme conditions on the thermodynamic properties of quark matter is one of the important applications of modern thermal quantum field theory. These conditions include high temperatures up to $10^{12}$ K, large densities up to $10^{14}$ gr/cm$^{3}$, large magnetic fields up to $10^{20}$ Gau\ss, and large angular velocities up to $10^{22}$ s$^{-1}$. These conditions are partly realized in nature, e.g. in the early universe or the core of compact stars. The Quark-Gluon plasma (QGP) produced in relativistic heavy-ion collision (HIC) experiments at Relativistic Heavy Ion Collider (RHIC) and Large Hadron Collider (LHC) also exhibits these extreme conditions. The aim of these experiments is to recreate the conditions after the big bang in laboratories. Various international projects and intensive studies are in progress to understand the nature of the matter produced after these collisions and to overcome the deficiencies of standard computational methods in simulating quark matter under extreme conditions \cite{rajagopal2018, pisarski2022, aarts2023, hot-QCD, present2023,becattini2021, machine2023, davoudi2024}.
\par
Among the aforementioned extreme conditions, analyzing quark matter under rotation has attracted much attention in the past few years \cite{vilenkin1980,rotation,rot2,rot3,rot4,rot5,rot6,rot7,rot8,rot9,rot10,rot11, fukushima2015, fukushima2018, sadooghi2021, zamora2021, fukushima2022, chernodub2023-1, chernodub2023-2,ambrus2023,zamora2023, sadooghi2023,cao2023,fukushima2024}. Several important phenomena, e.g. chiral vortical effect \cite{cve}, are related to the presence of a uniform rotation in relativistic systems \cite{fukushima2018}.
When apart from rotation, these systems are subjected to a uniform magnetic field, an inverse magnetorotational effect occurs \cite{fukushima2015}, that particularly leads to a reduction of the temperature of the chiral phase transition \cite{rotation,rot2,rot3,rot4,rot5,rot6,rot7,rot8,rot9,rot10,rot11,sadooghi2021}. Recently, it has been shown that this effect is the main reason for excluding certain phases of quark matter in the interior of neutron stars under some specific circumstances \cite{sadooghi2023}.
The field-theoretical investigation of rotation is immensely simplified once it is assumed that the system under consideration is under a rigid rotation. Although this kind of rotation cannot be attained in the expanding QGP produced in HICs, however, all theoretical investigations of this problem are based on this assumption. The latter has several unexpected consequences:    \\
The effect of rigid rotation on the equation of state of gluodynamics is studied recently in  \cite{chernodub2023-1,chernodub2023-2}. Assuming sufficiently small angular velocities $\Omega$, the free energy is Taylor expanded in powers of $\Omega$ up to $\mathcal{O}(\Omega^{2})$. This expansion leads immediately to an angular momentum density that is proportional to $\Omega$. According to classical mechanics, the proportionality factor is the moment of inertia $I(T)$. By computing $I(T)$ using numerical simulation of lattice quantum chromodynamics, it is shown that it receives two different contributions. The competition between these two temperature dependent terms leads to a negative moment of inertia at a temperature below a certain supervortical temperature, $T_{s}$. According to these results, $T_{s}$ is given by $T_{s}\approx 1.5 T_{c}$, where $T_{c}$ is the confinement/deconfinement phase transition temperature. Assuming that the angular momentum is finite, a vanishing moment of inertia at $T_{s}$ leads to the phenomenon of supervorticity, characterized by very large angular velocity at this temperature \cite{chernodub2023-1}. Moreover, as gluons are spin-one bosons, another interesting effect, dubbed '`negative Barnett effect`' \cite{chernodub2023-2}, is supposed to occur at temperatures below $T_{s}$. In contrast to the ordinary Barnett effect, in the negative Barnett effect, the rotation polarizes spin negatively. This is only possible when the moment of inertia $I_{S}$ related to spin $\boldsymbol{S}$ is negative, and its absolute value is larger than the moment of inertia $I_{L}$, related to the angular momentum $\boldsymbol{L}$. Since the latter is always positive \cite{chernodub2023-2}, the total moment of inertia $I=I_{L}+I_{S}$ corresponding to the total angular momentum $\boldsymbol{J}=\boldsymbol{L}+\boldsymbol{S}$ becomes negative.
\par
In the present paper, we intend to answer whether a negative moment of inertia also arises in a spin-zero relativistic gas. We thus analyze the impact of a rigid rotation on a relativistic Bose gas. Massless and massive scalar fields under rigid rotation are previously studied in \cite{ambrus2023}. Here, the focus is on imaginary rotation \cite{fukushima2022,cao2023,fukushima2024}, which has application in numerical simulation of a rigidly rotating system on the lattice. It is shown that this procedure leads to the appearance of the fractal features of thermodynamics under imaginary rotation and ninionic deformation of statistics, leading to stable ghostlike excitations \cite{ambrus2023}. In \cite{zamora2023}, the chiral symmetry breaking/restoration in a Yukawa model is studied. The authors determine first the propagators of free bosons and fermions in a rotating medium using the Fock-Schwinger proper-time method \cite{zamora2021}. These propagators are then used to determine the thermodynamic potential of the rigidly rotating Yukawa gas.
\par
In the following sections, we first review the results presented in \cite{zamora2023} and determine the free propagator of a rigidly rotating Bose gas using the Fock-Schwinger proper-time method. To do this, we start with the Lagrangian density of a complex Klein-Gordon (CKG) field $\phi$. We use the imaginary time formalism to introduce the temperature $T$ and determine the free propagator of the Bose gas at finite $T$. Introducing the interaction term $\lambda(\phi^{\dagger}\phi)^{2}$ in the Lagrangian density, we then utilize this propagator to determine the thermodynamic potential of this gas in the zeroth and first perturbative expansion in the orders of the coupling constant $\lambda$. We present the results in an integral form and compare it with the corresponding thermodynamic potential in a nonrotating Bose gas. As expected, the angular velocity $\Omega$ plays the role of a chemical potential \cite{fukushima2015}. We then perform an appropriate high temperature expansion (HTE) and present the corresponding perturbative part of the thermodynamic potential in this approximation. Apart from these parts, we determine the nonperturbative ring contribution to the thermodynamic potential. The final result for the thermodynamic potential, including the zeroth and first perturbative corrections as well as the nonperturbative ring potential exhibits a summation over $\ell$, which arises from the solution of the CKG equation of motion in cylinder coordinate system.\footnote{The quantum number $\ell$ is the conjugate momentum of the azimuthal angle $\varphi$ in a cylindrical coordinate system. } In the second part of the paper, we perform this summation numerically. Here, we mainly focus on the thermodynamic properties of the relativistic Bose gas under rigid rotation. Using the thermodynamic potential, we first determine the pressure of this gas and study the impact of a rigid rotation on this pressure.  Using standard thermodynamic relations, we also determine the angular momentum and entropy densities, $j$ and $s$, the heat capacity $C_{V}$, the speed of sound $c_{s}^{2}$, and the moment of inertia $I$. Setting first $\ell=1$, we present analytical expressions for these quantities up to $\mathcal{O}(\Omega^{2})$. Plotting the moment of inertia in terms of the coupling $\alpha\equiv \lambda/\pi^{2}$, it turns out that $I$ becomes negative for certain coupling $\alpha_{s}$, dubbed `'supervortical coupling''. In Fig. \ref{fig1}, we schematically describe how a negative moment of inertia affects the rotation of a system. In a system with a positive (negative) moment of inertia, an applied angular momentum $\boldsymbol{J}$ leads to a rotation with an angular velocity $\boldsymbol{\Omega}$ parallel (antiparallel) to $\boldsymbol{J}$.
Finally, we perform the summation over $\ell$ numerically and explore the $T$ and $\Omega$ dependence of the above thermodynamic quantities. We show that at high temperatures and for large coupling constants, certain thermodynamic instabilities appear. They are particularly manifested by negative $I$ and $C_{V}$, as well as large $c_{s}$. For very large couplings, $c_{s}$ becomes superluminal at high temperatures.
\par
The organization of this paper is as follows: In Sec. \ref{sec2}, we solve the equation of motion of a free CKG field under rotation. The free bosonic propagator at zero and finite temperature is presented in \ref{sec2B}. In Sec. \ref{sec3}, we compute the thermodynamic potential of a rigidly rotating Bose gas in the zeroth [Sec. \ref{sec3A}] and first perturbative level [Sec. \ref{sec3C}], as well as the nonperturbative ring potential [Sec. \ref{sec3D}]. To this purpose, we compute the one-loop tadpole diagram of this model in \ref{sec3B}.  In Sec. \ref{sec4}, we determine the thermodynamic quantities of this Bose gas under rigid rotation and explore the $T$ and $\Omega$ dependence of these quantities first in the first nonvanishing term in $\Omega$ for $\ell=1$ [Sec. \ref{sec4A}] and then numerically for $\ell>1$ [Sec. \ref{sec4B}]. Section \ref{sec5} is devoted to our concluding remarks. In Appendix \ref{appA}, we present the analytical details leading to the free propagator of the free CKG model. In Appendices \ref{appB} and \ref{appC}, we perform an appropriate HTE and present the free thermodynamic potential and one-loop tadpole diagram in this approximation.

\section{Complex Klein-Gordon fields under rotation}\label{sec2}
\setcounter{equation}{0}
\subsection{The model}\label{sec2A}
We start with the action of a free CKG field in a curved space-time
\begin{eqnarray}\label{S1}
S_{0}=\int\limits{{{d}^{4}}}x\left(-\det \left(g_{\mu \nu }\right) \right)^{1/2}\mathcal{L}_{0},
\end{eqnarray}
with the Lagrangian density
\begin{eqnarray}\label{S2}
\mathcal{L}_{0}=g^{\mu\nu}\partial_{\mu}\phi^{\dagger} {{\partial }_{\nu }}\phi -{{m}^{2}}\phi^{\dagger}\phi.
\end{eqnarray}
To study the effect of a rigid rotation on a relativistic Bose gas described by \eqref{S1}, we introduce the metric
\begin{eqnarray}\label{S3}
	{{g}_{\mu \nu }}=\left( \begin{matrix}
		1-\left( {{x}^{2}}+{{y}^{2}} \right){{\Omega }^{2}} & y\Omega  & -x\Omega  & 0  \\
		y\Omega  & -1 & 0 & 0  \\
		-x\Omega  & 0 & -1 & 0  \\
		0 & 0 & 0 & -1  \\
	\end{matrix} \right),
\end{eqnarray}
where $\Omega$ is a constant angular velocity.  The assumed rotation around the $z$-direction leads to a cylindrical symmetry around this axis. The system is thus naturally described by a cylindrical coordinate system $x^{\mu}=(t,x,y,z)=(t,r\cos\varphi, r\sin\varphi,z)$, with $r$ the radial coordinate, $\varphi$ the azimuthal angle, and $z$ the height of the cylinder. Plugging $\mathcal{L}$ from \eqref{S2} into the Euler-Lagrange equation of motion
\begin{eqnarray}\label{S4}
{{\partial }_{\alpha }}\left( \frac{\partial \mathcal{L}_{0}}{\partial \left( {{\partial }_{\alpha }}\phi^{\dagger}  \right)} \right)-\frac{\partial \mathcal{L}_{0}}{\partial \phi^{\dagger} }=0,
\end{eqnarray}
we arrive at
\begin{eqnarray}\label{S5}
{{\partial }_{\alpha }}\left( {{g}^{\alpha \nu }}{{\partial }_{\nu }}\phi  \right)+{{m}^{2}}\phi =0,
\end{eqnarray}
with $g^{\mu\nu}$, the inverse of $g_{\mu\nu}$ from \eqref{S3}. Using $L_{z}\equiv -i\left(x\partial_{y}-y\partial_{x}\right)=-i\partial_{\varphi}$, the equation of motion of a rotating CKG field in a cylindrical coordinate system reads
\begin{eqnarray}\label{S6}
\big[ {{\left( i{{\partial }_{t}}+\Omega {{L}_{z}} \right)}^{2}}+\boldsymbol{\nabla}^{2}+\partial _{z}^{2}-{{m}^{2}} \big]\phi \left( x \right)=0,
\end{eqnarray}
with $\boldsymbol{\nabla}^{2}= \partial _{r }^{2}+\frac{1}{r}{{\partial }_{r}}+\frac{1}{{{r}^{2}}}\partial _{\varphi }^{2}$. To solve \eqref{S6}, we use the ansatz
\begin{eqnarray}\label{S7}
\phi_{\ell}\left(x,k\right)={{e}^{-iEt+i{{k}_{z}}z+i\ell\varphi }}\mathcal{R}_{\ell}(r),
\end{eqnarray}
where the radial part of $\phi_{\ell}\left(x,k\right)$, $\mathcal{R}_{\ell}(r)$ satisfies
\begin{eqnarray}\label{S8}
\left( \partial _{r}^{2}+\frac{1}{r}{{\partial }_{r}}-\frac{{{\ell }^{2}}}{{{r}^{2}}}+k_{\perp }^{2} \right)\mathcal{R}_{\ell}(r)=0,
\end{eqnarray}
with $k_{\perp}^{2}\equiv \tilde{E}^{2}-k_{z}^{2}-m^{2}$ and $\tilde{E}\equiv E+\ell\Omega$. Introducing $\rho\equiv rk_{\perp}$, we finally arrive at
\begin{eqnarray}\label{S9}
\left[ {{\rho }^{2}}\partial _{\rho }^{2}+\rho {{\partial }_{\rho }}+({{\rho }^{2}}-{{\ell }^{2}}) \right]\mathcal{R}_{\ell}(\rho)=0,
\end{eqnarray}
which is the Bessel differential equation leading to $\mathcal{R}_{\ell}(r)=J_{\ell}(k_{\perp}r)$, where $J_{\ell}(z)$ is the Bessel function. Plugging this result into \eqref{S7}, the solution of \eqref{S6} reads
\begin{eqnarray}\label{S10}
\phi_{\ell}\left(x,k\right)={{e}^{-iEt+i{{k}_{z}}z+i\ell\varphi }}J_{\ell}(k_{\perp}r).
\end{eqnarray}
\subsection{Free bosonic propagator at zero and finite temperature}\label{sec2B}
According to the Fock-Schwinger proper-time method, the free two-point Green's function $D_{0}(x,x')$ of a CKG field is given by \cite{zamora2021, zamora2023}
\begin{eqnarray}\label{S11}
D_{0}\left( x,x' \right)=-i\int_{-\infty }^{0}{d\tau \sum\limits_{\lambda }{\exp \left( -i\lambda \tau  \right){{\phi }_{\lambda }}\left( x \right)\phi _{\lambda }^{\dagger }\left( x' \right)}},\nonumber\\
\end{eqnarray}
where $\lambda$ and $\phi_\lambda$ are the energy eigenvalue and eigenfunction of the differential operator $\mathcal{D}(\partial_{x},x)$. They arise by solving the eigenvalue equation
\begin{eqnarray}\label{S12}
\mathcal{D}(\partial_{x},x)\phi_{\lambda}(x)=\lambda\phi_{\lambda}(x).
\end{eqnarray}
To show \eqref{S11}, one starts with the Green's function differential equation
\begin{eqnarray}\label{S13}
\mathcal{D}\left(\partial_{x},x\right)D_{0}\left(x,x'\right)=\delta^{4}\left(x-x'\right),
\end{eqnarray}
where $D_{0}(x,x')$ is represented as
\begin{eqnarray}\label{S14}
D_{0}(x,x')=-i\int_{-\infty}^{0} U(x,x';\tau)~d\tau.
\end{eqnarray}
Here, $\tau$ is the proper-time and $U(x,x';\tau)$ is the proper-time evolution operator which satisfies
\begin{eqnarray}\label{S15}
i\partial_{\tau}U(x,x';\tau)=\mathcal{D}\left(\partial_{x},x\right)U\left(x,x';\tau\right).
\end{eqnarray}
Using the boundary conditions
\begin{eqnarray}\label{S16}
\lim\limits_{\tau\to 0}\mathcal{U}\left( x,{x}';\tau  \right)={{\delta }^{4}}\left( x-{x}' \right),\quad
\lim\limits_{\tau\to\infty}\mathcal{U}\left( x,{x}';\tau  \right)=0,  \nonumber\\
\end{eqnarray}
the solution of \eqref{S15} reads
\begin{eqnarray}\label{S17}
\mathcal{U}\left(x,x';\tau\right)=e^{-i\tau H(\partial_{x},x)}\delta^{4}\left(x-x'\right).
\end{eqnarray}
This result leads to \eqref{S11} upon using \eqref{S13} and the completeness relation satisfied by $\phi_{\lambda}(x)$
\begin{eqnarray}\label{S18}
\sum\limits_{\lambda }{{{\phi}_{\lambda }}\left(x \right)\phi _{\lambda }^{\dagger }\left( x' \right)= {{\delta }^{4}}\left( x-{x}' \right)}.
\end{eqnarray}
For $\mathcal{D}=\left( i\partial_{t}+\Omega {L}_{z} \right)^{2}+\boldsymbol{\nabla}^{2}+\partial _{z}^{2}-m^{2}$ from \eqref{S6}, the two-point Green's function of a CKG field under rotation is given by inserting \eqref{S10} into \eqref{S11}, with $\lambda=\tilde{E}^{2}-k_{\perp}^{2}-k_{z}^{2}-m^{2}$ \cite{zamora2023},
\begin{widetext}
\begin{eqnarray}\label{S19}
D_{0}(x,x')=-i\int_{-\infty}^{0}d\tau\sum_{\ell=-\infty}^{+\infty}\int\frac{dEdk_{z} k_{\perp}dk_{\perp}}{(2\pi)^{3}}e^{-i\tau\left(\tilde{E}^{2}-k_{\perp}^{2}-k_{z}^{2}-m^{2}+i\epsilon\right)}e^{-iE(t-t')+ik_{z}(z-z')+i\ell(\varphi-\varphi')}J_{\ell}(k_{\perp}r)J_{\ell}(k_{\perp}r'). \nonumber\\
\end{eqnarray}
\end{widetext}
Integrating \eqref{S19} over $\tau$ and performing a change of variable $E\to E-\ell\Omega$, we arrive at $D_{0}(x,x')$ in coordinate space
\begin{eqnarray}\label{S20}
D_{0}\left( x,x' \right)&=&\sum\limits_{\ell=-\infty}^{+\infty}
\int\frac{dEdk_zk_{\perp}dk_{\perp }}{\left( 2\pi  \right)^{3}}J_{\ell}\left(k_{\perp }r \right)J_{\ell}\left(k_{\perp }r' \right)\nonumber\\
&&\times\frac{e^{-iE\left( t-t' \right)+i\ell \Omega \left( t-t' \right)+ik_{z}\left( z-z' \right)+i\ell \left( \varphi -\varphi' \right)}}{E^{2}-k_{\perp }^{2}-k_{z}^{2}-m^{2}+i\epsilon}.\nonumber\\
\end{eqnarray}
The corresponding free propagator in the Fourier space is determined by
\begin{eqnarray}\label{S21}
D_{\ell\ell'}^{(0)}(p,p')=\int d^{4}xd^{4}x'D_{0}(x,x')\phi_{\ell}(x,p)\phi_{\ell'}(x',p'), \nonumber\\
\end{eqnarray}
with $d^{4}x=dtd\varphi dzrdr$ in the cylindrical coordinate system, $D_{0}(x,x')$ from \eqref{S21}, and $\phi_{\ell}(x,p)$ given in \eqref{S10}. Performing the integration over $x$ and $x'$, we arrive after some computation at the free boson propagator at zero temperature (see Appendix \ref{appA} for more details)
\begin{eqnarray}\label{S22}
D_{\ell\ell'}^{(0)}(p,p')=(2\pi)^{3}\widehat{\delta}^{3}_{\ell,\ell'}(p_{0},p_{z},p_{\perp};p^{\prime}_{0},p^{\prime}_{z},p^{\prime}_{\perp})D_{\ell}^{(0)}(p),\nonumber\\
\end{eqnarray}
with
\begin{eqnarray}\label{S23}
\widehat{\delta}^{3}_{\ell,\ell'}(p_{0},p_{z},p_{\perp};p^{\prime}_{0},p^{\prime}_{z},p^{\prime}_{\perp})&\equiv&\frac{1}{p_{\perp}}\delta\left(p_{0}-p^{\prime}_{0}\right)\delta\left(p_{z}-p^{\prime}_{z}\right)\nonumber\\
&&\times
\delta\left(p_{\perp}-p^{\prime}_{\perp}\right)\delta_{\ell\ell'},
\end{eqnarray}
and
\begin{eqnarray}\label{S24}
D_{\ell}^{(0)}(p_{0},\omega)\equiv\frac{1}{\left(p_{0}+\ell\Omega\right)^{2}-\omega^{2}+i\epsilon}.
\end{eqnarray}
Here, $\omega^{2}\equiv p_{\perp}^{2}+p_{z}^{2}+m^{2}$. At finite temperature $T$, $p_{0}$ is to be replaced with $i\omega_{n}$, where  $\omega_{n}=2\pi nT$ is the Matsubara frequency.
In the next section, we use
\begin{eqnarray}\label{S25}
D_{\ell}^{(0)}(\omega_{n},\omega)\equiv \frac{1}{\left(\omega_{n}-i\ell\Omega\right)^{2}+\omega^{2}},
\end{eqnarray}
to derive the thermodynamic potential of an interacting relativistic Bose gas under rotation up to first order in perturbative expansion. We also determine the nonperturbative ring potential in the lowest order.
\section{Thermodynamic potential of an interacting CKG field in the presence of rotation}\label{sec3}
\setcounter{equation}{0}
In this section, we determine the thermodynamic potential of an interacting CKG field in the presence of rotation. We start with the Lagrangian density
\begin{eqnarray}\label{A1}
\mathcal{L}=\mathcal{L}_{0}+\mathcal{L}_{\text{int}},
\end{eqnarray}
where the free part of the Lagrangian $\mathcal{L}_{0}$ is given in \eqref{S2}, and the interaction part reads
\begin{eqnarray}\label{A2}
\mathcal{L}_{\text{int}}=-\lambda \left(\phi^{\dagger}\phi\right)^{2}.
\end{eqnarray}
Here, $\lambda>0$ is the coupling constant of the model. Assuming that $\lambda< 1$, it is possible to perturbatively expand the thermodynamic potential $V_{\text{eff}}$ in a power series in the orders of $\lambda$,
\begin{eqnarray}\label{A3}
V_{\text{eff}}=\sum_{k=0}^{+\infty}\lambda^{k}V_{\text{eff}}^{(k)}.
\end{eqnarray}
In Sec. \ref{sec3A}, we first determine the exact expression of the zeroth order thermodynamic potential $V_{\text{eff}}^{(0)}$ by making use of the standard methods in thermal field theory \cite{kapusta-book, endrodi-book}. We then perform an appropriate HTE and present $V_{\text{eff}}^{(0)}$ in this approximation.
\par
To determine the one-loop contribution to the thermodynamic potential, $V_{\text{eff}}^{(1)}$, the one-loop self-energy function of the model, $\Pi_{1}$, is to be computed. In Sec. \ref{sec3B}, we first present an exact expression for $\Pi_{1}$. We then determine $\Pi_{1}$ in the limit of high temperature. We end this section by determining the exact expression of the nonperturbative ring potential $V_{\text{ring}}$ for this model.
\subsection{Zeroth order correction to the thermodynamic potential}\label{sec3A}
According to \cite{kapusta-book, lebellac-book}, the free (zeroth order correction) thermodynamic (effective) potential $V_{\text{eff}}$ of a relativistic Bose gas is given by
\begin{eqnarray}\label{A4}
V_{\text{eff}}^{(0)}&=&T\sum\limits_{n=-\infty}^{+\infty}\sum\limits_{\ell=-\infty}^{+\infty}\int\frac{dp_{z}p_{\perp}dp_{\perp }}{\left( 2\pi  \right)^{2}}\ln\left(\beta^{2}(D_{\ell}^{(0)})^{-1} \right),\nonumber\\
\end{eqnarray}
where $\beta\equiv T^{-1}$ and $D_{\ell}^{(0)}(\omega_{n},\omega)$ is the free propagator of this model.
Plugging \eqref{S25} into \eqref{A4}, we arrive after some standard computation at
\begin{eqnarray}\label{A5}
V_{\text{eff}}^{(0)}&=&\frac{T}{2}\sum\limits_{n=-\infty}^{+\infty}\sum\limits_{\ell=-\infty}^{+\infty}\sum_{\zeta=\pm}\int\frac{dp_{z}p_{\perp}dp_{\perp }}{\left( 2\pi  \right)^{2}}\nonumber\\
&&\times
\ln \left[\beta^{2}\left( \omega _{n}^{2}+\left( \omega +\zeta\ell \Omega  \right)^{2} \right) \right].
\end{eqnarray}
Using at this stage
\begin{eqnarray}\label{A6}
\hspace{-0.5cm}\sum\limits_{n=-\infty }^{+\infty }\ln \left(\left( 2n\pi  \right)^{2}+u^{2} \right)=u+2\ln \left( 1-e^{-u} \right),
\end{eqnarray}
we perform the summation over Matsubara frequencies. The zeroth order correction to $V_{\text{eff}}$ is thus given by
\begin{eqnarray}\label{A7}
\lefteqn{V_{\text{eff}}^{(0)}=\sum_{\ell=-\infty}^{+\infty}\int \frac{dp_{z}p_{\perp}dp_{\perp}}{(2\pi)^{2}}
\bigg\{\omega}\nonumber\\
&&+T\bigg[\ln \left( 1-e^{-\beta \left( \omega +\ell \Omega  \right)} \right)+\ln \left( 1-e^{-\beta \left( \omega -\ell \Omega  \right)} \right)\bigg]\bigg\}.\nonumber\\
\end{eqnarray}
The first term is the vacuum contribution to $V_{\text{eff}}^{(0)}$. It is independent of the angular velocity $\Omega$. The $T$-dependent part of $V_{\text{eff}}^{(0)}$, however, can be compared with the thermodynamic potential of a free relativistic Bose gas at finite chemical potential $\mu$ \cite{kapusta-book, laine-book,endrodi-book}. The fact that $\ell\Omega$ plays the role of the chemical potential $\mu$ is indeed expected from the literature (see, e.g.,  \cite{fukushima2015,sadooghi2021,sadooghi2023}). \par
In what follows, we present another possibility to evaluate \eqref{A4}. To do this, let us again start with \eqref{A4}. Using
\begin{eqnarray}\label{A8}
\ln a^{2}&=&-\frac{\partial}{\partial\kappa}\left(a^{2}\right)^{-\kappa}\bigg|_{\kappa=0},\nonumber\\
&=&-\frac{\partial}{\partial\kappa}\frac{1}{\Gamma(\kappa)}\int_{0}^{\infty}ds~s^{\kappa-1}e^{-a^{2}s}\bigg|_{\kappa=0},
\end{eqnarray}
we arrive first at
\begin{eqnarray}\label{A9}
\lefteqn{\hspace{-0.5cm}V_{\text{eff}}^{(0)}=-T\sum\limits_{n=-\infty }^{+\infty }\sum\limits_{\ell=-\infty }^{+\infty }\int\frac{dp_{z}p_{\perp}dp_{\perp}}{(2\pi )^{2}}\bigg[\frac{\partial }{\partial \kappa } \frac{1}{\Gamma \left( \kappa  \right)}
}\nonumber\\
&&\times \int_{0}^{\infty }ds\,s^{\kappa -1}e^{-\beta^{2}[(\omega_{n}-i\ell \Omega )^{2}+\omega^{2}]s} \bigg]\bigg|_{\kappa=0}.
\end{eqnarray}
Performing the summation over $n$, by making use of
\begin{equation}\label{A10}
\sum\limits_{n=-\infty }^{+\infty }e^{-\beta^{2}\left(\omega_{n}-i\ell \Omega  \right)^{2}s}=\frac{1}{2\sqrt{\pi s}}\vartheta_{3}\left(-\frac{i\ell\Omega\beta }{2}\bigg|\frac{i}{4\pi s} \right),
\end{equation}
where $\vartheta_{3}(z|\tau)$ is the elliptic theta-function \cite{gradstein},\footnote{Here, we have used the notation $\vartheta_{3}\left(z|\tau\right)\equiv \vartheta_{3}\left(z,e^{-i\pi\tau}\right)$. Here,  $\vartheta_{3}$ is given by
$$
\vartheta_{3}(z|\tau)=1+2\sum_{n=1}^{+\infty} \tau^{n^{2}}\cos(2n z).
$$}
 integrating over $p_{z}$ and $p_{\perp}$ according to \eqref{appB10}, and using
\begin{eqnarray}\label{A11}
\frac{d}{d\kappa}\left(\frac{s^{\kappa}}{\Gamma(\kappa)}\right)\bigg|_{\kappa=0}=1,
\end{eqnarray}
we obtain
\begin{eqnarray}\label{A12}
V_{\text{eff}}^{(0)}=-\frac{T^{4}}{16\pi^{2}}\sum\limits_{\ell =-\infty }^{+\infty}\int_{0}^{\infty }\frac{ds}{s^{3}}e^{-(m\beta)^{2}s}\vartheta_{3}\left( -\frac{i\ell\Omega\beta }{2}\bigg| \frac{i}{4\pi s}\right).\nonumber\\
\end{eqnarray}
For our numerical purposes, it is necessary to subtract the $T=0$ contribution from $V_{\text{eff}}^{(0)}$. We thus arrive at
\begin{eqnarray}\label{A13}
V_{\text{eff}}^{(0)T}=-\frac{T^{4}}{16\pi^{2}}\sum\limits_{\ell =-\infty }^{+\infty }\mathcal{A}_{3,\ell}(x,y),
\end{eqnarray}
where $\mathcal{A}_{3,\ell}(x,y)$ is given from
\begin{eqnarray}\label{A14}
\mathcal{A}_{n,\ell}(x,y)\equiv \int_{0}^{\infty}\frac{ds}{s^{n}}e^{-x^2 s}\left[\vartheta_{3}\left( \left. \frac{-i\ell y}{2} \right| \frac{i}{4\pi s} \right) -1\right], \nonumber\\
\end{eqnarray}
by choosing $n=3$. Here, $x\equiv m\beta$ and $y\equiv \Omega\beta$.
\par
In Sec. \ref{sec4}, we study the thermodynamic properties of a relativistic Bose gas under rotation by making use of \eqref{A13}. We derive the pressure, the entropy density, the angular momentum, and the energy density up to first order perturbative corrections inclusive the first corrections to the nonperturbative ring potential.
\par
Inspired by the method presented in \cite{laine-book,toms-book}, it is possible to expand $V_{\text{eff}}^{(0)}$ in $x\ll 1$ and $y\ll 1$ and to determine an approximation of this potential at high temperature.
According to the proof presented in Appendix \ref{appB}, the HTE of $V_{\text{eff}}^{(0)T}$ from \eqref{A7} reads
\begin{eqnarray}\label{A15}
V_{\text{eff}}^{(0) T}&=&-T^{4}
\left\{\frac{\pi^{2}}{45}-\frac{x^{2}}{12}+\frac{x^{3}}{6\pi }
\right.\nonumber\\
&&\left.-\frac{x^{4}}{16{{\pi }^{2}}}\left( \ln \left( \frac{4\pi}{x} \right)-{{\gamma }_{\text{E}}}+\frac{3}{4} \right)\right.\nonumber\\
&&\left. +\sum\limits_{\ell =1}^{+\infty}\left(\frac{3x^{2}-\ell^{2}y^{2}}{12\pi^{2}}-\frac{x}{2\pi}+\frac{1}{3}\right)\ell^{2}y^{2}\right\}+\cdots. \nonumber\\
\end{eqnarray}
Similar expression appears also in \cite{kapusta-book,weldon-haber}, where $\ell y=\ell\Omega\beta$ is replaced with the chemical potential $\mu$. According to this result, for $m,\Omega=0$, we thus have
\begin{eqnarray}\label{A16}
V_{\text{eff}}^{(0)T}\xrightarrow{m,\Omega\to 0}-\frac{\pi^{2}T^{4}}{45},
\end{eqnarray}
which is the thermodynamic potential of a free and massless relativistic Bose gas \cite{kapusta-book}. Since we are interested in the $\Omega$ corrections to $V_{\text{eff}}^{(0)T}$, it is possible to keep the first nonvanishing $\Omega$ dependent term in \eqref{A15}.
For $\ell=1$, we thus have
\begin{eqnarray}\label{A17}
V_{\text{eff}}^{(0)T}\approx -\frac{\pi^2T^{4}}{45}\left(1+\frac{15}{\pi^{2}}(\Omega\beta)^{2}\right).
\end{eqnarray}
This result indicates that rotation increases the pressure of a free relativistic Bose gas. In Sec. \ref{sec4}, we study the effect of $\Omega\beta$ on the pressure of free relativistic Bose gas arising from \eqref{A13} and show that this statement is true also once $\ell>1$ contributions are taken into account. We also
compare it with the pressure arising from \eqref{A15} in the high temperature limit. We show that at a certain temperature these two expressions coincides.
\subsection{One-loop perturbative correction to the self-energy function }\label{sec3B}
\begin{figure}[hbt]
\includegraphics[width=12cm, height=3.5cm]{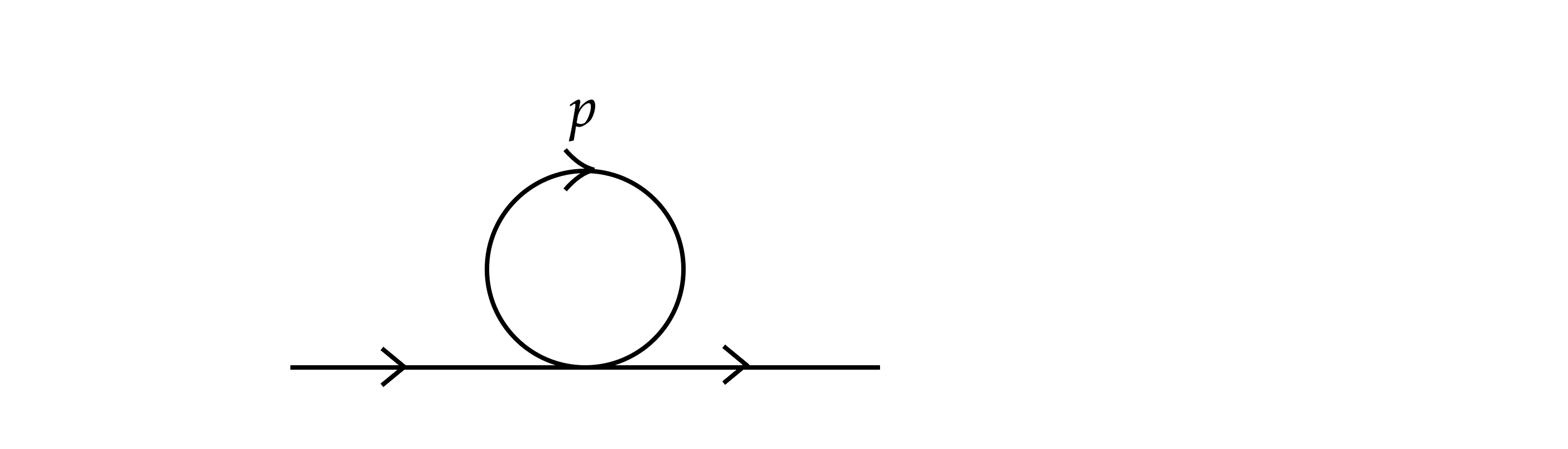}
\vspace{0cm}
\caption{One-loop self-energy diagram $\Pi_{1}$. }\label{fig2}
\end{figure}

The one-loop correction to the self-energy function, $\Pi_{1}$, is given by the tadpole diagram from Fig. \ref{fig2}. Using the free propagator $D_{\ell}^{(0)}(\omega_{n},\omega)$ from \eqref{S25}, it is  given by
\begin{eqnarray}\label{A18}
\Pi_{1}=4\lambda T\sum_{n=-\infty}^{+\infty}\sum_{\ell=-\infty}^{+\infty}\int\frac{dp_{z}p_{\perp}dp_{\perp}}{(2\pi)^{2}}D_{\ell}^{(0)}(\omega_{n},\omega),\nonumber\\
\end{eqnarray}
with $D_{\ell}^{(0)}(\omega_{n},\omega)$ from \eqref{S25}. It is possible to utilize the method presented in the previous section and determine $\Pi_{1}$ in an exact form. To do this, we use
\begin{eqnarray}\label{A19}
D_{\ell}^{(0)}(\omega_{n},\omega)=\frac{1}{2\omega}\frac{\partial}{\partial\omega}\ln\left(\beta^{2}\left(D_{\ell}^{(0)}\right)^{-1}\right),
\end{eqnarray}
and replace the propagator in \eqref{A18} with the expression on the right-hand side of  \eqref{A19}.
According to the method leading from \eqref{A4} to \eqref{A13}, we arrive first at
\begin{eqnarray}\label{A20}
\lefteqn{\Pi_{1}=-2\lambda T\sum_{n=-\infty}^{+\infty}\sum_{\ell=-\infty}^{+\infty}\int\frac{dp_{z}p_{\perp}dp_{\perp}}{(2\pi)^{2}}\frac{1}{\omega}
}\nonumber\\
&&\times \frac{\partial}{\partial\omega}\bigg[\frac{\partial}{\partial\kappa}\frac{1}{\Gamma(\kappa)}\int_{0}^{\infty}ds~s^{\kappa-1}e^{-\beta^{2}[(\omega_{n}-i\ell\Omega)^{2}+\omega^{2}]s}\bigg]\bigg|_{\kappa=0}.\nonumber\\
\end{eqnarray}
After performing the summation over the Matsubara frequencies by making use of \eqref{A10}, integrating over $p_{z}$ and $p_{\perp}$ according to \eqref{appB10}, and using \eqref{A11} as well as
\begin{eqnarray}\label{A21}
\frac{1}{\omega}\frac{\partial}{\partial\omega}\left(e^{-\beta^{2}\omega^{2}s}\right)=-2s\beta^{2}e^{-\beta^{2}\omega^{2}s},
\end{eqnarray}
we arrive at the temperature dependent part of $\Pi_{1}$
\begin{eqnarray}\label{A22}
\Pi_{1}^{\text{mat}}=\frac{\alpha T^{2}}{4}\sum_{\ell=-\infty}^{+\infty}\mathcal{A}_{2,\ell}(x,y),
\end{eqnarray}
where $\alpha\equiv \lambda/\pi^{2}$ and $\mathcal{A}_{2,\ell}(x,y)$ can be read from \eqref{A14}.
\par
In what follows, we perform a HTE and present the matter part of $\Pi_{1}$ in this approximation. To do this, let us consider \eqref{A18} and evaluate the summation over the Matsubara frequencies by making use of
\begin{eqnarray}\label{A23}
\lefteqn{\hspace{-1cm}\sum\limits_{n=-\infty }^{+\infty }{\frac{1}{{{\left( {{\omega }_{n}}-i\ell \Omega  \right)}^{2}}+{{\omega }^{2}}}}
}\nonumber\\
&=&\frac{\beta }{2\omega }\left(n_{b}\left( \omega +\ell \Omega  \right)+n_{b}\left( \omega -\ell \Omega  \right)+1 \right),
\end{eqnarray}
where $n_{b}$ is the Bose-Einstein distribution function defined by
\begin{eqnarray}\label{A24}
n_{b}(\omega)\equiv\frac{1}{e^{\beta\omega}-1}.
\end{eqnarray}
Plugging \eqref{A23} into \eqref{A18}, we arrive at
\begin{eqnarray}\label{A25}
\Pi_{1}=\Pi_{1}^{\text{vac}}+\Pi_{1}^{\text{mat}},
\end{eqnarray}
with the vacuum ($T=0$) part
\begin{eqnarray}\label{A26}
\Pi_{1}^{\text{vac}}\equiv 2\lambda\sum_{\ell=-\infty}^{+\infty}\int\frac{dp_{z}p_{\perp}dp_{\perp}}{(2\pi)^{2}}\frac{1}{\omega},
\end{eqnarray}
and the matter ($T\neq 0$) part
\begin{eqnarray}\label{A27}
\Pi_{1}^{\text{mat}}&\equiv& 2\lambda\sum\limits_{\ell=-\infty}^{+\infty}\int\frac{dp_{z}p_{\perp}dp_{\perp}}{(2\pi)^{2}}\frac{1}{\omega}\nonumber\\
&&\times [n_{b}\left(\omega+\ell\Omega\right)+n_{b}\left(\omega-\ell\Omega\right)].
\end{eqnarray}
We focus only on $\Pi_{1}^{\text{mat}}$ and separate $\ell=0$ and $\ell\neq 0$ contribution of $\Pi_{1}^{\text{mat}}$ to obtain
\begin{eqnarray}\label{A28}
\Pi_{1}^{\text{mat}}=4\lambda\left(\mathcal{J}_{1}+\mathcal{J}_{2}\right),
\end{eqnarray}
with
\begin{eqnarray}\label{A29}
\mathcal{J}_{1}&\equiv& \int\frac{dp_{z}p_{\perp}dp_{\perp}}{(2\pi)^{2}}\frac{n_{b}(\omega)}{\omega},\nonumber\\
\mathcal{J}_{2}&\equiv&\sum_{\ell=1}^{+\infty}\int\frac{dp_{z}p_{\perp}dp_{\perp}}{(2\pi)^{2}}\frac{1}{\omega} [n_{b}\left(\omega+\ell\Omega\right)+n_{b}\left(\omega-\ell\Omega\right)]. \nonumber\\
\end{eqnarray}
In Appendix \ref{appC}, we apply the method used in Appendix  \ref{appB} and perform a HTE of $\mathcal{J}_{i}, i=1,2$ from \eqref{A29}. The resulting expressions are presented in \eqref{appC7} and \eqref{appC13}. Combining these expression the matter part of $\Pi_{1}$ for $x\ll 1$ and $y\ll 1$ is given by
\begin{eqnarray}\label{A30}
\lefteqn{\hspace{-1cm}\Pi_{1}^{\text{mat}}=4\lambda T^{2}\left\{\frac{1}{12}-\frac{x}{4\pi }+ \right.\frac{x^{2}}{8\pi^{2}}\left( \ln \left( \frac{4\pi}{x} \right)-\gamma_{\text{E}}+\frac{1}{2} \right) }\nonumber\\
	 &&+ \left.\sum\limits_{\ell =1}^{+\infty}\left[\frac{1}{6}-
\frac{x}{2\pi}\left(1-\frac{\ell^{2}y^{2}}{2x^{2}}\right)
-\frac{\ell^{2}y^{2}}{4\pi ^{2}}\right.\right.\nonumber\\
&&\left.\left.+\frac{x^{2}}{4\pi^{2}}\left( \ln \left( \frac{4\pi}{x} \right)-\gamma_{\text{E}}+\frac{1}{2} \right) \right] \right\}+\cdots.
\end{eqnarray}
In the limit of vanishing $m$ and $\Omega$, $\Pi_{1}^{\text{mat}}$ is given by\footnote{For $\Omega=0$, we neglect the series over $\ell$ in \eqref{A30}.}
\begin{eqnarray}\label{A31}
\Pi_{1}^{\text{mat}}\xrightarrow{m,\Omega\to 0}\Pi_{0}\equiv \frac{\lambda T^{2}}{3}.
\end{eqnarray}
Apart from a factor, this result is, as expected, the same as the one presented in \cite{kapusta-book} for the one-loop self-energy diagram of a $\lambda\varphi^{4}$ theory. Let us remind that $\lambda T^{2}$ plays the role of a thermal mass for charged bosons.
\par
Taking the limit of $m\beta\to 0$ in \eqref{A30}, and keeping the first nonvanishing term in $\Omega$, we arrive for $\ell=1$ at
\begin{eqnarray}\label{A32}
\Pi_{1}^{\text{mat}}\approx \lambda T^{2}\left(1-\frac{(\Omega\beta)^{2}}{\pi^{2}}\right).
\end{eqnarray}
To arrive at \eqref{A32}, we particularly used $\Omega<m$ and neglected $\left(\Omega\beta/m\beta\right)^{2}$ in the second line of \eqref{A30}.
This result indicates that, at least in the limit of $m\beta\to 0$, the rotation decreases the thermal mass of a charged boson.
\subsection{One-loop perturbative correction to the thermodynamic potential }\label{sec3C}
Following the arguments in \cite{kapusta-book}, the one-loop contribution to the thermodynamic potential is given by
\begin{eqnarray}\label{A33}
V_{\text{eff}}^{(1)}=\lambda\left(T\sum\limits_{n=-\infty}^{+\infty}\sum\limits_{\ell=-\infty}^{+\infty}\int\frac{dp_{z}p_{\perp}dp_{\perp}}{(2\pi)^{2}}D_{\ell}^{(0)}(\omega_{n},\omega)\right)^{2}.\nonumber\\
\end{eqnarray}
Comparing \eqref{A33} with \eqref{A18} and neglecting the $T$-independent part of the thermodynamic potential, it is possible to determine $V_{\text{eff}}^{(1)T}$ using the one-loop self-energy function $\Pi_{1}^{\text{mat}}$,
\begin{eqnarray}\label{A34}
V_{\text{eff}}^{(1)}=\frac{1}{16\lambda}\left(\Pi_{1}^{\text{mat}}\right)^{2}.
\end{eqnarray}
The exact expression for $\Pi_{1}^{\text{mat}}$ is given in \eqref{A22} and its HTE is presented in \eqref{A30}.
Using \eqref{A22}, we thus obtain
\begin{eqnarray}\label{A35}
V_{\text{eff}}^{(1)T}=\frac{\alpha T^{4}}{256\pi^{2}}\left(\sum_{\ell=-\infty}^{+\infty}\mathcal{A}_{2,\ell}(x,y)\right)^{2}.
\end{eqnarray}
In the high temperature limit,
\begin{eqnarray}\label{A36}
\lefteqn{\hspace{-1cm}V_{\text{eff}}^{(1)T}=\lambda T^{4}\left\{\frac{1}{12}-\frac{x}{4\pi }+ \right.\frac{x^{2}}{8\pi^{2}}\left( \ln \left( \frac{4\pi}{x} \right)-\gamma_{\text{E}}+\frac{1}{2} \right) }\nonumber\\
	 &&+ \left.\sum\limits_{\ell =1}^{+\infty}\left[\frac{1}{6}-
\frac{x}{2\pi}\left(1-\frac{\ell^{2}y^{2}}{2x^{2}}\right)
-\frac{\ell^{2}y^{2}}{4\pi ^{2}}\right.\right.\nonumber\\
&&\left.\left.+\frac{x^{2}}{4\pi^{2}}\left( \ln \left( \frac{4\pi}{x} \right)-\gamma_{\text{E}}+\frac{1}{2} \right) \right] \right\}^{2}+\cdots,
\end{eqnarray}
arises from \eqref{A30}. The thermodynamic potential up to one-loop perturbative correction is thus given by
\begin{eqnarray}\label{A37}
V_{\text{eff}}^{T}=V_{\text{eff}}^{(0)T}+V_{\text{eff}}^{(1)T},
\end{eqnarray}
with $V_{\text{eff}}^{(0)T}$ from \eqref{A13} or \eqref{A15} and $V_{\text{eff}}^{(1)T}$ from \eqref{A35} or \eqref{A36}. \par
In the high temperature limit $x\ll 1$ and $y\ll 1$, it is possible to neglect the $m$ and $\Omega$ dependent terms in \eqref{A36}. The $T$ dependent part of $V_{\text{eff}}^{(1)T}$ is thus given by
\begin{eqnarray}\label{A38}
V_{\text{eff}}^{(1)T}\xrightarrow{m,\Omega\to 0}\frac{\lambda T^{4}}{144}.
\end{eqnarray}
Together with $V_{\text{eff}}^{(0)}\approx -\frac{\pi^{2}T^{4}}{45}$ from \eqref{A16}, we arrive at $V_{\text{eff}}^{T}$ in this approximation,
\begin{eqnarray}\label{A39}
V_{\text{eff}}^{T}\xrightarrow{m,\Omega\to 0} -\frac{\pi^{2}T^{4}}{45}\left(1-\frac{45}{144}\alpha\right).
\end{eqnarray}
Keeping the first nonvanishing term in $\Omega$ in \eqref{A36}, we arrive for $\ell=1$ at
\begin{eqnarray}\label{A40}
V_{\text{eff}}^{(1)T}\approx \frac{\lambda T^{4}}{16}\left(1-\frac{2(\Omega\beta)^{2}}{\pi^{2}}\right),
\end{eqnarray}
that together with the zeroth order correction to $V_{\text{eff}}$ from \eqref{A17} leads to
\begin{eqnarray}\label{A41}
V_{\text{eff}}^{T}\approx -\frac{\pi^{2}T^{4}}{45}\bigg[\left(1-\frac{45}{16}\alpha\right)+\frac{15(\Omega\beta)^{2}}{\pi^{2}}\left(1+\frac{3}{8}\alpha\right)\bigg].\nonumber\\
\end{eqnarray}
In what follows, we determine the nonperturbative ring contribution to the thermodynamic potential.
\subsection{Nonperturbative ring contribution and the total thermodynamic potential}\label{sec3D}
\begin{figure}[hbt]
\includegraphics[width=11cm, height=5cm]{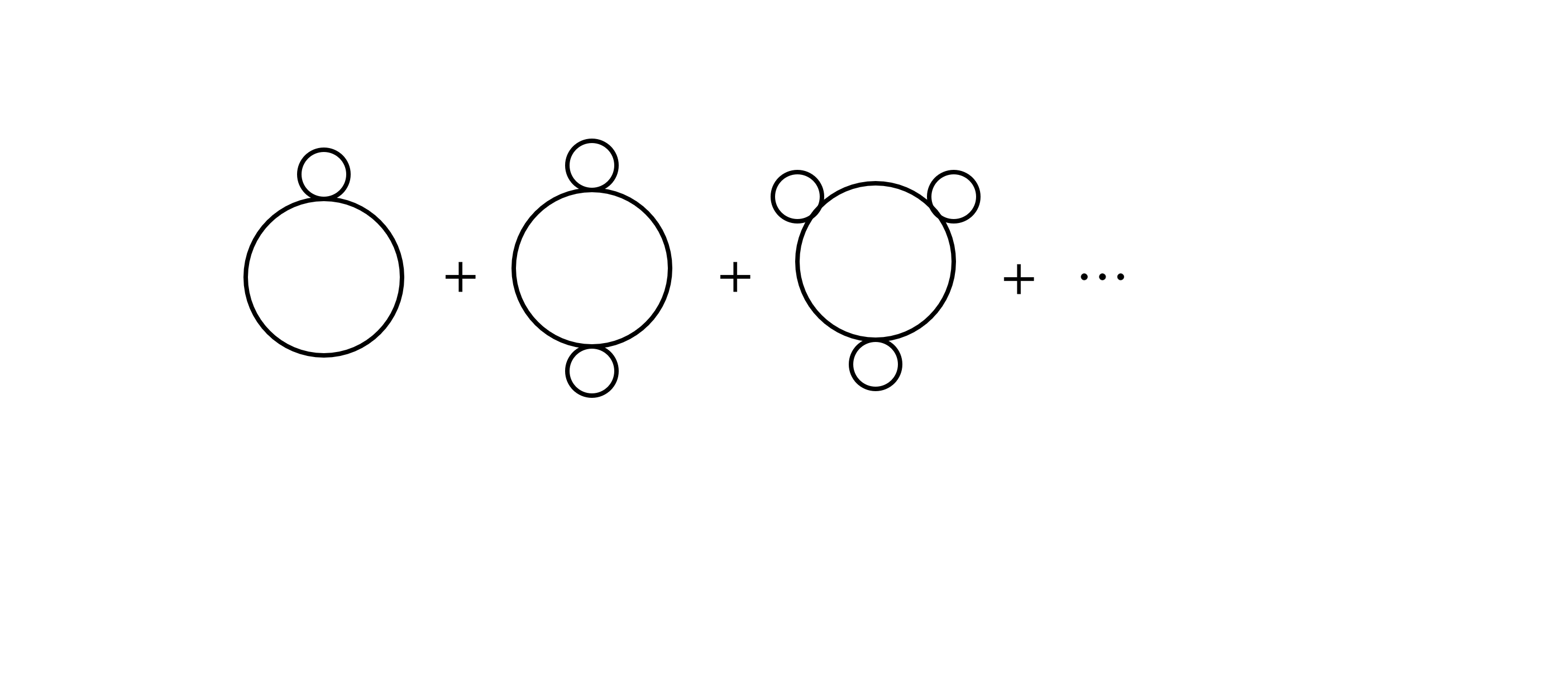}
\vspace{-2cm}
\caption{The ring diagrams of an interacting CKG model. Small circles indicate the $\Pi_{1}$-insertion. }\label{fig3}
\end{figure}

Following the arguments in \cite{kapusta-book}, the nonperturbative part of the thermodynamic potential is given by the ring potential
\begin{eqnarray}\label{A42}
\lefteqn{\hspace{-0.5cm}V_{\text{ring}}=T\sum\limits_{n=-\infty}^{+\infty}\sum\limits_{\ell=-\infty}^{+\infty}\int\frac{dp_{z}p_{\perp}dp_{\perp}}{(2\pi)^{2}}
}\nonumber\\
&&\times\bigg[\ln\left(1+\Pi_{\ell} D_{\ell}^{(0)}\right)-\Pi_{\ell} D_{\ell}^{(0)}\bigg],
\end{eqnarray}
which arises by the resummation of ring diagrams with an increasing number of $\Pi_{1}$-insertion (see Fig. \ref{fig3}).
\vspace{-1cm}
Here,
$$
\Pi_{\ell}\equiv 4\lambda T\sum_{n=-\infty}^{+\infty}\int\frac{dp_{z}p_{\perp}dp_{\perp}}{(2\pi)^{2}}D_{\ell}^{(0)}(\omega_{n},\omega),\nonumber\\
$$
arises from $\Pi_{1}=\sum_{\ell=-\infty}^{+\infty}\Pi_{\ell}$ with the one-loop self-energy diagram $\Pi_{1}$ from \eqref{A18}\footnote{Later, we consider only the $T$-dependent part of $\Pi_{\ell}$ in $V_{\text{ring}}$.} and $D_{\ell}^{(0)}$ the free boson propagator from \eqref{S25}. In what follows, we determine the leading contribution to $V_{\text{ring}}$ by considering $n=0$ in the summation over the Matsubara frequencies. Plugging $D_{\ell}^{(0)}$ with $n=0$ into \eqref{A42}, we first obtain
\begin{eqnarray}\label{A43}
V_{\text{ring}}&=&T\sum\limits_{\ell=-\infty}^{+\infty}\int\frac{dp_{z}p_{\perp}dp_{\perp}}{(2\pi)^{2}}
\nonumber\\
&&\times\bigg\{\ln\left(1+\frac{\Pi_{\ell}}{[p_{z}^{2}+p_{\perp}^{2}+m^2-(\ell\Omega)^{2}]}\right)\nonumber\\
&&\qquad-\frac{\Pi_{\ell}}{[p_{z}^{2}+p_{\perp}^{2}+m^2-(\ell\Omega)^{2}]}\bigg\}.
\end{eqnarray}
Inspired by the method described in Appendices \ref{appB} and \ref{appC}, it is possible to perform the integration over $p_{z}$ and $p_{\perp}$ and arrive at an exact expression for $V_{\text{ring}}$.
To do this, we replace the logarithm in \eqref{A43} with its Taylor series,
\begin{eqnarray}\label{A44}
\ln\left(1+x\right)=\sum_{k=1}^{+\infty}\frac{(-1)^{k+1}x^{k}}{k},
\end{eqnarray}
and arrive at
\begin{eqnarray}\label{A45}
V_{\text{ring}}&=&T\sum\limits_{\ell=-\infty}^{+\infty}\int\frac{dp_{z}p_{\perp}dp_{\perp}}{(2\pi)^{2}}\sum_{k=2}^{+\infty}\frac{(-1)^{k+1}}{k}(\Pi_{\ell})^{k}\left(u^{2}\right)^{-k},\nonumber\\
\end{eqnarray}
where $u^{2}\equiv p_{z}^{2}+p_{\perp}^{2}+m^2-(\ell\Omega)^{2}$. Using \eqref{appB11}, we have
\begin{eqnarray}\label{A46}
(u^{2})^{-k}=\frac{1}{\Gamma(k)}\int_{0}^{\infty}dt t^{k-1}e^{-u^{2}t}.
\end{eqnarray}
Plugging this expression into \eqref{A45} and performing the integration over $p_{z}$ and $p_{\perp}$ by using \eqref{appB10}, we obtain
\begin{eqnarray}\label{A47}
V_{\text{ring}}&=&\frac{\pi^{3/2}T}{(2\pi)^{3}}\sum\limits_{\ell=-\infty}^{+\infty}\sum\limits_{k=2}^{+\infty}\frac{(-1)^{k+1}}{k\Gamma(k)}(\Pi_{\ell})^{k}\nonumber\\
&&\times\int_{0}^{\infty}dt t^{k-5/2}e^{-\zeta_{\ell} t},
\end{eqnarray}
with $\zeta_{\ell}\equiv [m^2-(\ell\Omega)^{2}]$. Assuming that $\text{Re}[\zeta_{\ell}]>0$, it is possible to perform the integration over $t$ according to\footnote{In the massless limit, it is useful to first replace $\Omega$ with $i\Omega_{I}$ and eventually analytically continue back to $\Omega$. In this way $\zeta_{\ell}$ becomes positive and  integration over $t$ in \eqref{A48} will be possible.}
\begin{eqnarray}\label{A48}
\int_{0}^{\infty}dt~t^{k-5/2}e^{-\zeta_{\ell} t}=\Gamma\left(k-3/2\right)\zeta_{\ell}^{3/2-k}.
\end{eqnarray}
Substituting this expression into \eqref{A47} and performing the summation over $k$, we arrive at
\begin{eqnarray}\label{A49}
\lefteqn{\hspace{-0.5cm}\sum\limits_{k=2}^{+\infty}\frac{(-1)^{k+1}}{k}\frac{\Gamma(k-3/2)}{\Gamma(k)}\left(\Pi_{\ell}\right)^{k}\zeta_{\ell}^{3/2-k}}\nonumber\\
&&=\frac{2\pi^{1/2}}{3} \bigg[3\zeta_{\ell}^{1/2}\Pi_{\ell} -2\left(\Pi_{\ell} +\zeta_{\ell}\right)^{3/2}+2 \zeta_{\ell}^{3/2}\bigg].\nonumber\\
\end{eqnarray}
Plugging finally \eqref{A49} into \eqref{A47}, the ring potential \eqref{A47} reads
\begin{eqnarray}\label{A50}
V_{\text{ring}}&=&\frac{T}{12\pi}\sum_{\ell=-\infty}^{+\infty}\left(3\zeta_{\ell}^{1/2}\Pi_{\ell} -2\left(\Pi_{\ell} +\zeta_{\ell}\right)^{3/2}+2 \zeta_{\ell}^{3/2}\right).\nonumber\\
\end{eqnarray}
Adding $V_{\text{ring}}$ from \eqref{A50} to $V_{\text{eff}}^{T}$ from \eqref{A37}, the full thermodynamic potential up to one-loop perturbative correction inclusive the nonperturbative ring potential is thus given by
\begin{eqnarray}\label{A51}
V_{\text{eff}}=V_{\text{eff}}^{(0)T}+V_{\text{eff}}^{(1)T}+V_{\text{ring}}.
\end{eqnarray}
In Sec. \ref{sec4}, we use \eqref{A51} to study the thermodynamic behavior of a relativistic Bose gas under rigid rotation. In the rest of this section, we focus on $V_{\text{ring}}$ and determine it in the following four special cases:
\par
\textit{i) Case 1:} Let us first consider the massless limit.  Setting $m=0$ in $\zeta_{\ell}$, plugging the resulting expression into \eqref{A50}, and neglecting the terms with odd powers in $\ell$ in the summation over $\ell$, we are left with
\begin{eqnarray}\label{A52}
V_{\text{ring}}&=&-\frac{T}{6\pi}\sum_{\ell=-\infty}^{+\infty}\left(\Pi_{\ell} -(\ell\Omega)^{2}\right)^{3/2}.
\end{eqnarray}
Plugging $\Pi_{0}\equiv\lambda T^{2}/3$ from \eqref{A31} into \eqref{A52} and assuming that the Bose gas does not rotate, we arrive at
\begin{eqnarray}\label{A53}
V_{\text{ring}}\xrightarrow{m,\Omega\to 0}-\frac{\lambda^{3/2}T^{4}}{18\sqrt{3}},
\end{eqnarray}
as expected. Plugging this expression together with \eqref{A39} into \eqref{A51}, the full thermodynamic potential in the limit $m,\Omega\to 0$ reads
\begin{eqnarray}\label{A54}
\hspace{-1cm}V_{\text{eff}}\xrightarrow{m,\Omega\to 0} -\frac{\pi^{2}T^{4}}{45}\left(1-\frac{45}{144}\alpha+\frac{15}{6\sqrt{3}}\alpha^{3/2}\right).
\end{eqnarray}
This results is similar to the one presented in \cite{kapusta-book} for nonrotating relativistic neutral Bose gas.
\par
\textit{ii) Case 2:} To keep the lowest $\Omega$-dependent contribution to $V_{\text{ring}}$ in the massless limit, we replace $\Pi_{\ell}$ in \eqref{A52} with $\Pi_{1}^{\text{mat}}$ from \eqref{A32}. Here, only the $\ell=1$ term in the HTE of $\Pi_{1}^{\text{mat}}$ from \eqref{A30} is considered. Going through the same procedure leading to \eqref{A54}, we arrive first at
\begin{eqnarray}\label{A55}
\lefteqn{\hspace{-0.5cm}V_{\text{ring}}\approx -\frac{\lambda^{3/2}T^{4}}{6\pi}\bigg[\left(1-\frac{(\Omega\beta)^{2}}{\pi^{2}}\right)^{3/2}
}\nonumber\\
&&+2\sum\limits_{\ell=1}^{+\infty}\left(1-\frac{(\Omega\beta)^{2}}{\pi^{2}}-\frac{\ell^{2}(\Omega\beta)^{2}}{\lambda}\right)^{3/2}\bigg].
\end{eqnarray}
Considering only the contribution from $\ell=0,1$ terms, we obtain
\begin{eqnarray}\label{A56}
\hspace{-1cm}V_{\text{ring}}\approx-\frac{\lambda^{1/2}T^{4}}{2\pi}\left\{\lambda-(\Omega\beta)^{2}\left(1+\frac{3\lambda}{2\pi}\right)\right\}.
\end{eqnarray}
The full thermodynamic potential, including the perturbative part $V_{\text{eff}}^{T}$ from \eqref{A41} and the nonperturbative part $V_{\text{ring}}$ from \eqref{A56} in the massless limit is thus given by
\begin{eqnarray}\label{A57}
V_{\text{eff}}&\approx&-T^{4}\left(\mathcal{C}_0+\left(\Omega\beta\right)^{2}\mathcal{C}_{2}\right),
\end{eqnarray}
with  $T$ and $\Omega$ independent coefficients $\mathcal{C}_0$ and $\mathcal{C}_{2}$
\begin{eqnarray}\label{A58}
\mathcal{C}_0&\equiv&
\frac{\pi^{2}}{45}
\left(
1-\frac{45}{16}\alpha+\frac{45}{2}\alpha^{3/2}
\right),\nonumber\\
\mathcal{C}_{2}&\equiv&\frac{1}{3}\left(1-\frac{3}{2}\alpha^{1/2}+\frac{3}{8}\alpha-\frac{9}{4}\alpha^{3/2}\right).
\end{eqnarray}
A comparison between \eqref{A57} with \eqref{A54} shows that the limit $\Omega\to 0$ is singular.
\par
\textit{iii) Case 3:} In this case, we keep the $m$-dependence in $\zeta_{\ell}=m^{2}-(\ell\Omega)^{2}$ in \eqref{A50} and neglect the $m,\Omega$-dependence in $\Pi_{\ell}$. We thus arrive at
\begin{eqnarray}\label{A59}
V_{\text{ring}}&=&\frac{T}{12\pi}\sum_{\ell=-\infty}^{+\infty}\bigg\{
3\zeta_{\ell}^{1/2}\Pi_{0}
-2\left(\Pi_{0}+\zeta_{\ell}\right)^{3/2}+2\zeta_{\ell}^{3/2}\bigg\},\nonumber\\
\end{eqnarray}
with $\Pi_{0}$ from \eqref{A31}.
\par
\textit{iv) Case 4:} In this case, similar to the previous one, we keep the $m$-dependence in $\zeta_{\ell}$, and replace $\Pi_{\ell}$ in \eqref{A50} with
\begin{eqnarray}\label{A60}
\Pi_{\ell}(T,\Omega)\equiv\lambda\left(\frac{T^{2}}{3}-\frac{\ell^{2}\Omega^{2}}{2\pi^{2}}\right),
\end{eqnarray}
from \eqref{A30}. We thus obtain
\begin{eqnarray}\label{A61}
V_{\text{ring}}&=&\frac{T}{12\pi}\sum_{\ell=-\infty}^{+\infty}\bigg\{
3\zeta_{\ell}^{1/2}\Pi_{\ell}
-2\left(\Pi_{\ell}+\zeta_{\ell}\right)^{3/2}+2\zeta_{\ell}^{3/2}\bigg\},\nonumber\\
\end{eqnarray}
with $\Pi_{\ell}$ from \eqref{A60}. In Secs. \ref{sec4A} and \eqref{sec4B}, the above results are used to determine the thermodynamic quantities of a rigidly rotating relativistic Bose gas.
\section{Thermodynamic quantities of a rigidly rotating relativistic Bose gas}\label{sec4}
\setcounter{equation}{0}
In this section, we compute the pressure $P$, the angular momentum, entropy, and energy densities $j,s$, and $\epsilon$ by making use of the results from previous section. We also determine the moment of inertia $I$ and the heat capacity of this Bose gas analytically as well as numerically and show that in some regions in the parameter space of this model, they become negative.
Let us first consider the thermodynamic Euler equation of this system, $\epsilon+P=Ts$. Here, the pressure $P$ is given by
\begin{eqnarray}\label{M1}
P=-V_{\text{eff}},
\end{eqnarray}
with $V_{\text{eff}}$, the full thermodynamic potential from \eqref{A51}. It includes contributions from tree level, first perturbative correction as well as nonperturbative ring potential. The energy density $\epsilon$ is defined in the corotating frame. Its relation with the energy density in the nonrotating frame, $\epsilon^{nr}$, is given by $\epsilon=\epsilon^{nr}-j\Omega$. Here, $j$ is the angular momentum density of the rotating system. It is defined by
\begin{eqnarray}\label{M2}
 j\equiv\left(\frac{dP}{d\Omega}\right)_{T}.
\end{eqnarray}
This expression arises from the Gibbs-Duhem relation \cite{chernodub2023-1,chernodub2023-2}
\begin{eqnarray}\label{M3}
dP=sdT+jd\Omega.
\end{eqnarray}
Using \eqref{M3}, the entropy density $s$ is defined by
\begin{eqnarray}\label{M4}
s\equiv\left(\frac{dP}{dT}\right)_{\Omega}.
\end{eqnarray}
Apart from these quantities, let us define the heat capacity by \cite{kapusta-book}
\begin{eqnarray}\label{M5}
C_{V}\equiv\frac{d^{2}P}{dT^{2}}=\frac{ds}{dT},
\end{eqnarray}
and the speed of sound $c_{s}$,
\begin{eqnarray}\label{M6}
c_{s}^{2}\equiv\frac{dP}{d\epsilon}=\frac{s}{TC_{V}}.
\end{eqnarray}
We also define the moment of inertia $I=I(T)$ by Taylor expanding the pressure $P(T,\Omega)$ in the powers of $\Omega$,
\begin{eqnarray}\label{M7}
P(T,\Omega)=\sum_{n=0}^{+\infty}\frac{1}{n!}P^{(n)}(T,0)\Omega^{n},
\end{eqnarray}
with $P^{(n)}(T,0)\equiv\lim\limits_{\Omega\to 0}\frac{d^{n}P(T,\Omega)}{d\Omega^{n}}$, and identifying $P^{(2)}(T,0)$ with $I(T)$ \cite{chernodub2023-1,chernodub2023-2},
\begin{eqnarray}\label{M8}
I(T)\equiv \frac{d^{2}P(T,\Omega)}{d\Omega^{2}}\bigg|_{\Omega=0}.
\end{eqnarray}
Using \eqref{M2}, the moment of inertia \eqref{M8} can be also interpreted as the linear response to $j$,
\begin{eqnarray}\label{M9}
I(T)= \frac{j(T,\Omega)}{\Omega}\bigg|_{\Omega\to 0}.
\end{eqnarray}
Plugging \eqref{A13}, \eqref{A35}, and  \eqref{A50} with $\Pi_{\ell}$ from \eqref{A22} into $V_{\text{eff}}$ from \eqref{A51} the exact expression for the pressure $P$ in \eqref{M1} is determined. We use this exact result in Sec. \ref{sec4B} to study the thermodynamic properties of a relativistic Bose gas under a rigid rotation. In the following Sec. \ref{sec4A}, however, we present analytical results for $P$, $s$, $j$, $I$, and $\epsilon$ using the approximations \eqref{A54} (\textit{case i}) and \eqref{A57} (\textit{case ii}) for $V_{\text{eff}}$.
\subsection{Analytical results including first nontrivial contribution from $\boldsymbol{\ell}=1$}\label{sec4A}
Plugging $V_{\text{eff}}$ from \eqref{A54} (\textit{case i}) into \eqref{M1}, the pressure $P$ in the limit of vanishing $m$ and $\Omega$ reads
\begin{eqnarray}\label{M10}
\hspace{-0.5cm}P\xrightarrow{m,\Omega\to 0} \frac{\pi^{2}T^{4}}{45}\left(1-\frac{45}{144}\alpha+\frac{15}{6\sqrt{3}}\alpha^{3/2}\right).
\end{eqnarray}
This result is in analogy to the result presented in \cite{kapusta-book} for the pressure of an interacting relativistic neutral Bose gas ($\lambda\varphi^{4}$-theory). The nonanalytical contribution proportional to $\alpha^{3/2}$ arises from the nonperturbative ring contribution.
\par
\begin{figure}[hbt]
\includegraphics[width=8cm, height=5cm]{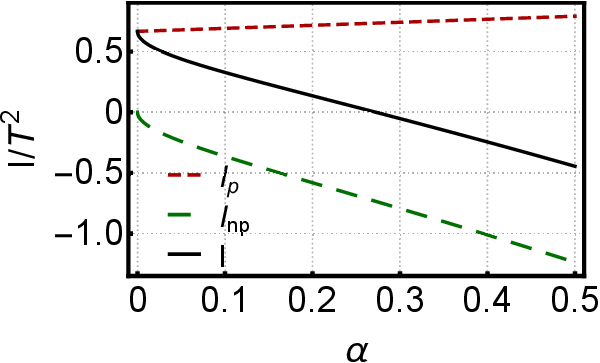}
\caption{The $\alpha$ dependence of $I_{\text{p}}, I_{\text{np}}$, and $I=I_{\text{p}}+I_{\text{np}}$ is demonstrated. Whereas $I_{\text{p}}>0$ and $I_{\text{np}}<0$ in the interval $\alpha\in [0,0.5]$, $I$ is positive for $\alpha<0.272$ and negative for $\alpha>0.272$.}\label{fig4}
\end{figure}

Plugging in contrast \eqref{A57} (\textit{case ii}) into \eqref{M1}, the first nontrivial contribution of $\Omega$ arises in the pressure as
\begin{eqnarray}\label{M11}
P\approx T^{4}\left(\mathcal{C}_0+\left(\Omega\beta\right)^{2}\mathcal{C}_{2}\right),
\end{eqnarray}
with $\mathcal{C}_{i}, i=0,2$ defined in \eqref{A58}. As in the previous case, in the coefficients $\mathcal{C}_{i},i=0,2$ the terms proportional to $\alpha^{n}$ and $\alpha^{n/2}$ with $n\in \mathbb{N}_{0}$ arise from the perturbative and ring corrections to the pressure $P$, respectively.  Using \eqref{M11}, we immediately arrive at $j,s, C_{V},c_{s}^{2}$, and $I$ in this approximation,
\begin{eqnarray}\label{M12}
j&\approx& 2T^2\mathcal{C}_{2}\Omega,\nonumber\\
s&\approx& 2T^{3}\left(2\mathcal{C}_{0}+\left(\Omega\beta\right)^{2}\mathcal{C}_{2}\right),\nonumber\\
C_{V}&\approx&2T^{2}\left(6\mathcal{C}_{0}+\left(\Omega\beta\right)^{2}\mathcal{C}_{2}\right),\nonumber\\
c_{s}^{2}&\approx& \frac{2\mathcal{C}_{0}+\left(\Omega\beta\right)^{2}\mathcal{C}_{2}}{6\mathcal{C}_{0}+\left(\Omega\beta\right)^{2}\mathcal{C}_{2}}\approx \frac{1}{3}\left(1+\frac{\mathcal{C}_{2}}{3\mathcal{C}_{0}}(\Omega\beta)^{2}\right),\nonumber\\
I&\approx& 2T^2\mathcal{C}_{2}.
\end{eqnarray}
Moreover, plugging \eqref{M11} and \eqref{M12} into $\epsilon=-P+Ts$, the energy density of the relativistic Bose gas reads
\begin{eqnarray}\label{M13}
\epsilon\approx T^{4}\left(3\mathcal{C}_{0}+(\Omega \beta)^2\mathcal{C}_{2}\right).
\end{eqnarray}
Let us emphasize that the coefficients $\mathcal{C}_{i},i=0,2$ depend only on $\alpha$. The thermodynamic quantities $j,s,C_V,c_{s}^{2}$ thus depend on $\alpha$ and $\Omega\beta$. The moment of inertia, however, depends only on $\alpha$. It consists of a perturbative and a nonperturbative part, $I=I_{\text{p}}+I_{\text{np}}$, with
\begin{eqnarray}\label{M14}
I_{\text{p}}(\alpha)&\equiv&\frac{2T^{2}}{3}\left(1+\frac{3}{8}\alpha\right), \nonumber\\
I_{\text{np}}(\alpha)&\equiv&-T^{2}\left(\alpha^{1/2}+\frac{3}{2}\alpha^{3/2}\right).
\end{eqnarray}
In Fig. \ref{fig4}, the $\alpha$ dependence of $I_{\text{p}},I_{\text{np}}$, and $I$ is demonstrated. It is shown that for $0<\alpha<0.5$, $I_{p}>0$ (red dashed line) and $I_{np}<0$ (green dashed line) in this interval, so that their combination $I$ becomes positive for $\alpha<0.272$ and negative for $\alpha>0.272$ (see black solid line).  At  $\alpha\approx 0.272$, the total moment of inertia vanishes. This scenario is very similar to the one described in \cite{chernodub2023-1, chernodub2023-2} in which the summation of two different contributions to $I$ leads to negative moment of inertia in some region of the parameter space.
Inspired by \cite{chernodub2023-1, chernodub2023-2}, we refer to $\alpha\approx 0.272$ at which the moment of inertia vanishes as supervortical coupling, $\alpha_{s}$.
According to \eqref{M12}, at this point $\mathcal{C}_{2}$ vanishes, thus, the speed of sound $c_{s}^{2}$ becomes equal to the speed of sound of a free relativistic Bose gas $c_{s}^{2}=1/3$. \\
Let us notice that for a perturbative expansion to be valid, $\alpha=\lambda/\pi^{2}$  must be lower than $0.1$. Thus assuming that $\alpha<0.1$, the total moment of inertia turns out to be always positive. In what follows, we determine numerically the thermodynamic quantities $j,s,C_V,c_s^{2},I$, and $\epsilon$ and show that by considering the contribution from $\ell>1$, some thermodynamic quantities, in particular, the moment of inertia and the heat capacity becomes negative for $\alpha<0.1$.
\subsection{Numerical results including contributions from $\boldsymbol{|\ell|\geq 0}$}\label{sec4B}
\subsubsection{Preliminaries}\label{sec4B1}
As it is shown in the previous section, the pressure $P$ includes three different contribution, the zeroth order correction $P_{0}$, the one-loop perturbative correction $P_{1}$, and the nonperturbative ring correction $P_{\text{ring}}$. There are given by $P_{0}=-V_{\text{eff}}^{(0)T}, P_{1}=-V_{\text{eff}}^{(1)T}$, and $P_{\text{ring}}=-V_{\text{ring}}$ with $V_{\text{eff}}^{(0)T},V_{\text{eff}}^{(1)T}, $ and $V_{\text{ring}}$ given in \eqref{A13}, \eqref{A35}, and \eqref{A50}. As concerns the ring contribution, we present in this section the results arising from \eqref{A59} (\textit{case iii}), where we use the lowest order contribution to the one-loop self-energy in \eqref{A50}.\footnote{We have also performed the computation with \eqref{A61} from \textit{case iv}. The difference between the results of \textit{case iii} and \textit{case iv} is negligible. } Since, according to \eqref{M2}-\eqref{M9}, other thermodynamic quantities arise from $P$, they also consists of three contribution $\mathcal{X}_{0}, \mathcal{X}_{1}$ and $\mathcal{X}_{\text{ring}}$ with $\mathcal{X}=\{j,s,C_V,c_{s}^{2}, I, \epsilon\}$. In what follows, we present the necessary analytic expressions for  $\mathcal{Y}_{0}, \mathcal{Y}_{1}$ and $\mathcal{Y}_{\text{ring}}$ with $\mathcal{Y}=\{\bar{j},\bar{s},\bar{I},\bar{C}_{V}\}$, where $\bar{j}\equiv j/T^{3}, \bar{s}\equiv s/T^{2},\bar{I}\equiv I/T^{2}$, and $\bar{C}_{V}\equiv C_{V}/T^{2}$ are dimensionless quantities. The zeroth order, one-loop, and ring contributions of $c_s^{2},$ and $\epsilon$ arise simply from these expressions. Using the analytical expressions in this section, we explore the $z\equiv T/m$ as well as $y=\Omega\beta$ behavior of these quantities. We particularly focus on the interval $z\in [0.1,1]$ and $y\in[0.01,0.025]$ as well as $\alpha\in[0.01,0.1]$. For nonvanishing $\Omega$, we numerically perform the summation over $\ell$ up to $\ell_{\text{max}}=50$.\footnote{We considered various upper limits for $\ell$ smaller and lower than $\ell_{\text{max}}=50$. It turns out that when $\ell_{\text{max}}\gtrsim 50$, the results remain stable and are qualitatively the same as those reported in the following sections. This robustness led us to confidently choose $\ell_{\text{max}}=50$. It is important to note that choosing $\ell_{\text{max}}\gg 50$ is not allowed due to the properties of $\vartheta_{3}(z|\tau)$ appearing in our analytical results.}
\par
Let us start with $\bar{j}^{(0)}$ arising from \eqref{M2} with $P$ replaced with $P_{0}$. Using \eqref{A13}, we arrive at
\begin{eqnarray}\label{M15}
\bar{j}_{0}(x,y) =-\frac{i}{32\pi^2}\sum_{\ell=-\infty}^{+\infty}\ell \mathcal{A}^{(1)}_{3,\ell}(x,y),
\end{eqnarray}
where $\mathcal{A}^{(m)}_{n,\ell}(x,y)$ is defined by
\begin{eqnarray}\label{M16}
\mathcal{A}_{n,\ell}^{(m)}(x,y)\equiv\int_{0}^{\infty}\frac{ds}{s^{n}}e^{-x^2 s}\vartheta^{(m)}_{3}\left( \left. \frac{-i\ell y}{2} \right| \frac{i}{4\pi s} \right).\nonumber\\
\end{eqnarray}
Here, $\vartheta^{(m)}_{3}(z|\tau)\equiv\frac{d^{m}}{dz^{m}}\vartheta_{3}(z|\tau)$. Plugging then $P_{1}$ into \eqref{M2} and using
$
\frac{dP_{1}}{d\Omega}=\beta\frac{dP_{1}}{dy}$, as well as
\begin{eqnarray}\label{M17}
\hspace{-0.5cm}\frac{d^{m}\mathcal{A}_{n,\ell}(x,0)}{dy^{m}}=\left(\frac{-i\ell}{2}\right)^{m}\mathcal{A}_{n,\ell}^{(m)}(x,y)\bigg|_{y=0},
\end{eqnarray}
 we arrive at
\begin{eqnarray}\label{M18}
\hspace{-0.8cm}\bar{j}_{1}(x,y,\alpha)=\frac{i\alpha}{256\pi^{2}}\sum_{\ell=-\infty}^{+\infty}\mathcal{A}_{2,\ell}\sum_{\ell=-\infty}^{+\infty}\ell\mathcal{A}_{2,\ell}^{(1)}.
\end{eqnarray}
Here, $\mathcal{A}_{2,\ell}$ and $\mathcal{A}_{2,\ell}^{(1)}$ are given by \eqref{A14} and \eqref{M16}, respectively. As concerns $\bar{j}_{\text{ring}}$, we use \eqref{A59} (\textit{case i}) and obtain
\begin{eqnarray}\label{M19}
\bar{j}_{\text{ring}}(x,y,\alpha)&=&\frac{y}{4\pi}\sum_{\ell=-\infty}^{+\infty}\ell^{2}\bigg[
\bar{\zeta}_{\ell}^{-1/2}\bar{\Pi}_{0}-2\left(\bar{\Pi}_{0}+\bar{\zeta}_{\ell}\right)^{1/2}\nonumber\\
&&+2\bar{\zeta}_{\ell}^{1/2}\bigg],
\end{eqnarray}
where $\bar{\zeta}_{\ell}\equiv x^{2}-\ell^{2}y^{2}$ and $\bar{\Pi}_{0}=\alpha\pi^{2}/3$ are dimensionless quantities.
\par
The zeroth order contribution to entropy density arises from \eqref{M4} with $P$ replaced with $P_{0}$. Using
\begin{eqnarray}\label{M20}
T\frac{d\mathcal{A}_{n,\ell}}{dT}&=&2x^{2}\mathcal{A}_{n-1,\ell}+\frac{i\ell y}{2}\mathcal{A}_{n,\ell}^{(1)},
\end{eqnarray}
we obtain
\begin{eqnarray}\label{M21}
\bar{s}_{0}(x,y)&=&\frac{1}{16\pi^{2}}\sum_{\ell=-\infty}^{+\infty}\left(4 \mathcal{A}_{3,\ell}+ 2 x^{2} \mathcal{A}_{2,\ell}+ \frac{i\ell y}{2} \mathcal{A}_{3,\ell}^{(1)}\right).\nonumber\\
\end{eqnarray}
To determine the one-loop contribution to the entropy density, we substitute $P_{1}$ into \eqref{M4}, to arrive first at
\begin{eqnarray*}
\bar{s}_{1}=-\frac{\alpha}{64\pi^{2}}\sum_{\ell=-\infty}^{+\infty}\mathcal{A}_{2,\ell}\sum_{\ell=-\infty}^{+\infty}\left(\mathcal{A}_{2,\ell}+\frac{T}{2}\frac{d\mathcal{A}_{2,\ell}}{dT}\right).
\end{eqnarray*}
Then using \eqref{M20}, we obtain
\begin{eqnarray}\label{M22}
\bar{s}_{1}(x,y,\alpha)&=&-\frac{\alpha}{64\pi^{2}}\sum_{\ell=-\infty}^{+\infty}\mathcal{A}_{2,\ell}\sum_{\ell=-\infty}^{+\infty}\bigg(\mathcal{A}_{2,\ell}+ x^{2} \mathcal{A}_{1,\ell}\nonumber\\
&&+ \frac{i\ell y}{4} \mathcal{A}_{2,\ell}^{(1)}\bigg).
\end{eqnarray}
Using \eqref{A59} and \eqref{M4}, the ring contribution to the entropy density reads
\begin{eqnarray}\label{M23}
\bar{s}_{\text{ring}}(x,y,\alpha)&=&-\frac{1}{6\pi}\sum_{\ell=-\infty}^{+\infty}\bigg[
\frac{9}{2}\bar{\zeta_{\ell}}^{1/2}\bar{\Pi}_{0}-\left(\bar{\Pi}_{0}+\bar{\xi}_{\ell}\right)^{3/2}\nonumber\\
&&+\bar{\zeta}_{\ell}^{3/2}-3\bar{\Pi}_{0}\left(\bar{\Pi}_{0}+\bar{\zeta}_{\ell}\right)^{1/2}
\bigg].
\end{eqnarray}
\begin{figure*}[hbt]
\includegraphics[width=8cm, height=6cm]{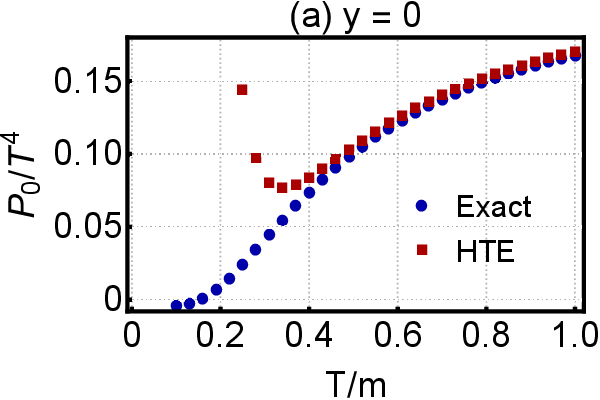}
\includegraphics[width=8cm, height=6cm]{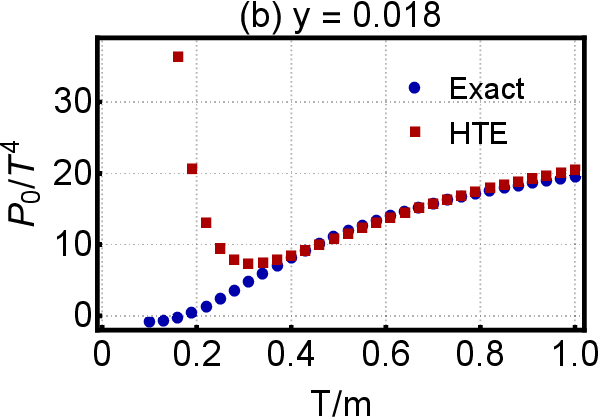}
\caption{A comparison between the $T/m$ dependence of the exact expression and the HTE of dimensionless $P_{0}/T^{4}$ are made for a nonrotating relativistic Bose gas with $y=0$ (panel a) and a rotating relativistic Bose gas with $y= 0.018$ (panel b). In both cases the HTE results coincide with the exact ones at $T/m\geq 0.4$. Moreover, the rotation increases the pressure $P_{0}$ up to several orders of magnitude.}\label{fig5}
\end{figure*}
\begin{figure*}[hbt]
\includegraphics[width=8cm, height=6cm]{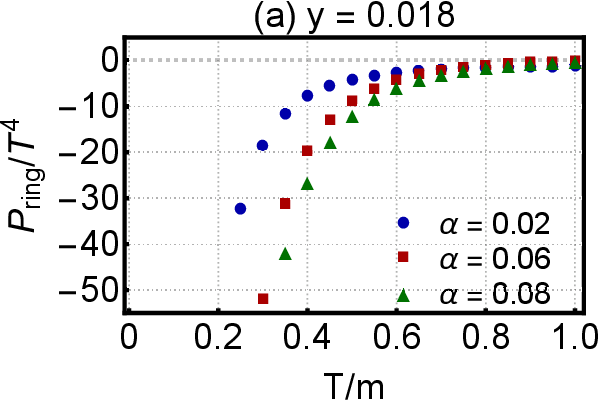}
\includegraphics[width=8cm, height=6cm]{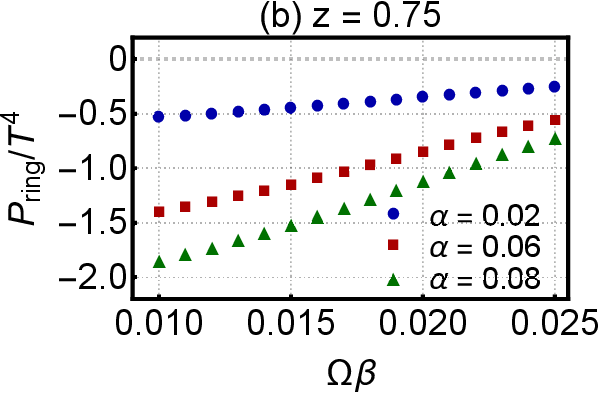}
\caption{The $T/m$ (panel a) and $\Omega\beta$ (panel b) dependence of dimensionless $P_{\text{ring}}/T^4$ are plotted for fixed $y=0.018$ (panel a) and $z=0.75$ (panel b) as well as $\alpha=0.02,0.06,0.08$. As it turns out, $P_{\text{ring}}$ is negative for all values of $0<\alpha<0.1$. For fixed angular velocity (temperature), it increases with increasing temperature (angular velocity). Its dependence on $\alpha$ is nontrivial (see Fig \ref{fig7}).}\label{fig6}
\end{figure*}
\begin{figure}[hbt]
\includegraphics[width=8cm, height=6cm]{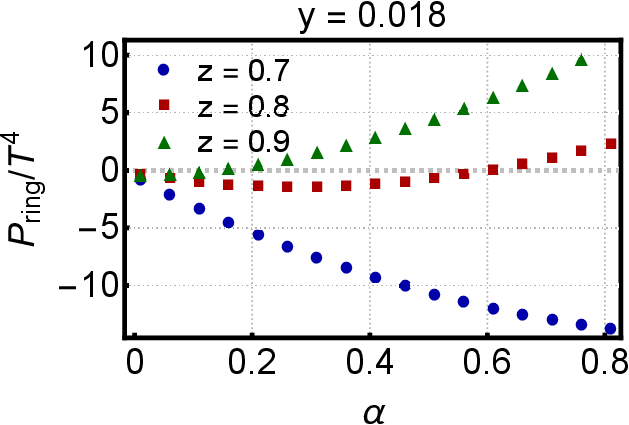}
\caption{The $\alpha$ dependence of dimensionless $P_{\text{ring}}/T^{4}$ is presented for fixed $y=0.018$ and $z=0.7, 0.8, 0.9$. Whereas for $z=0.7$, $P_{\text{ring}}/T^{4}$ is negative and decreases with increasing $\alpha$, for $z=0.8$ and $z=0.9$, it is first negative and then, after passing a minimum, increases with increasing $\alpha$. We notice that $\alpha>0.1$ corresponds to $\lambda>1$, which is not appropriate for perturbative studies.}\label{fig7}
\end{figure}
\begin{figure*}[hbt]
\includegraphics[width=8cm, height=6cm]{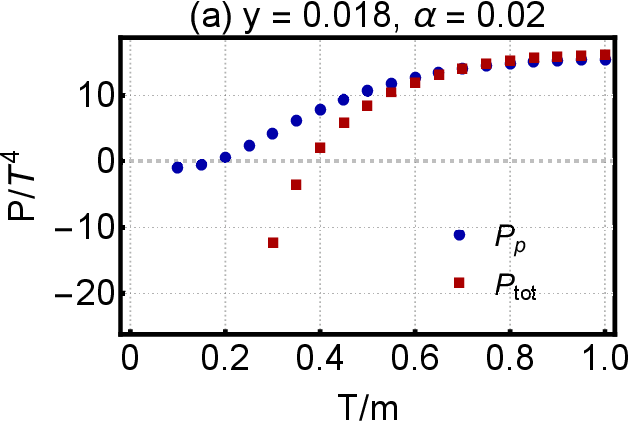}
\includegraphics[width=8cm, height=6cm]{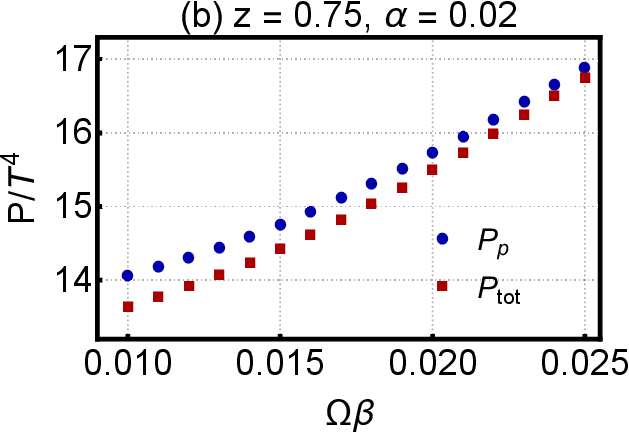}
\caption{The $T/m$ (panel a) and $\Omega\beta$ (panel b) dependence of dimensionless $P_{\text{p}}/T^4$ (blue circles) and $P_{\text{tot}}/T^{4}$ (red squares) are plotted for fixed $y=0.018$ (panel a) and $z=0.75$ (panel b) as well as $\alpha=0.02$. At high temperature, i.e. for $z>0.4$, $P_{\text{tot}}$ is positive and increases with increasing $T/m$ (see panel a). For fixed $z$, $P_{\text{tot}}$ increases with increasing $\Omega\beta$ (see panel b). The difference between $P_{\text{p}}$ and $P_{\text{tot}}$ becomes negligible at high temperature, as expected from Fig. \ref{fig6}.}\label{fig8}
\end{figure*}
\begin{figure*}[hbt]
\includegraphics[width=8cm, height=6cm]{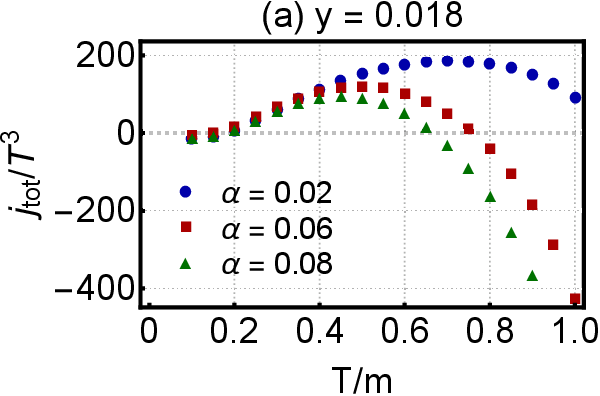}
\includegraphics[width=8cm, height=6cm]{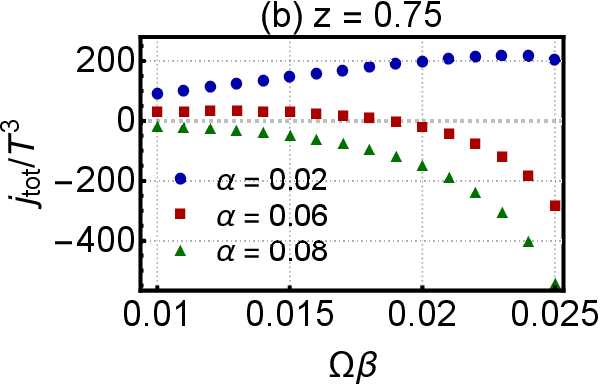}
\caption{The $T/m$ (panel a) and $\Omega\beta$ (panel b) dependence of dimensionless angular momentum density $j_{\text{tot}}/T^3$ are plotted for fixed $y=0.018$ (panel a) and $z=0.75$ (panel b) for $\alpha=0.02,0.06,0.08$. As it is demonstrated in panel a, $j_{\text{tot}}/T^{3}$ first increases  and then decreases with increasing $T/m$ and $\Omega\beta$. For larger values of $\alpha$, the temperature at which $j_{\text{tot}}/T^{3}$ vanishes is lower. Negative slopes of $j_{\text{tot}}/T^{3}$ indicates negative moment of inertia at high temperatures $T/m$ as well as high angular velocities $\Omega\beta$. The plot in panel b shows that $j_{\text{tot}}$ is not linear in $\Omega\beta$. }\label{fig9}
\end{figure*}
\begin{figure*}[hbt]
\includegraphics[width=8cm, height=6cm]{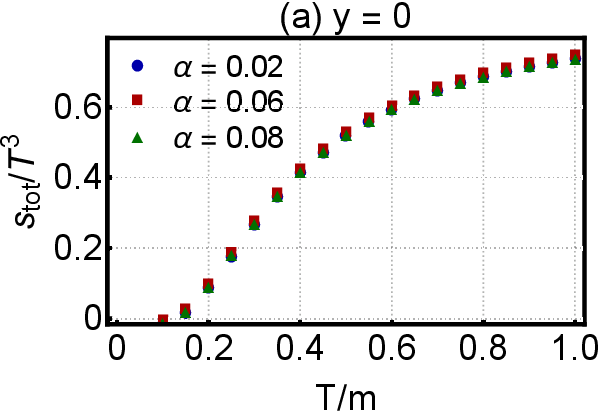}
\includegraphics[width=8cm, height=6cm]{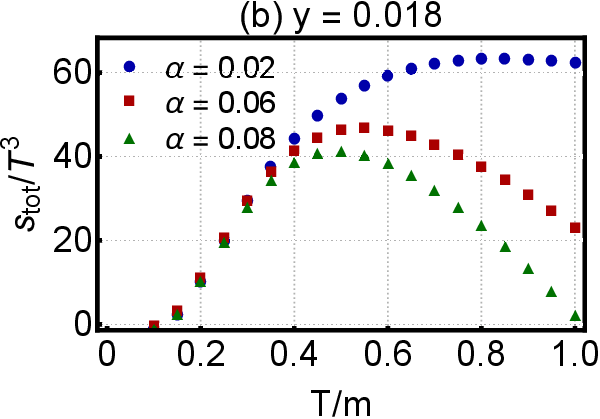}
\caption{(a) The $T/m$ dependence of dimensionless entropy density $s_{\text{tot}}/T^{3}$ in a nonrotating relativistic Bose gas is plotted for fixed $\alpha=0.02,0.06,0.08$. The entropy is increasing with increasing temperature. The coupling $\alpha$ has very small effect on this behavior, so that the results for different $\alpha$s almost coincide. (b) The $T/m$ dependence of dimensionless entropy density $s_{\text{tot}}/T^{3}$ in a rotating relativistic Bose gas is plotted for fixed $y=0.018$ and $\alpha=0.02,0.06,0.08$.
Whereas for a weakly interacting relativistic Bose gas with $\alpha=0.02$ the entropy density increases with increasing $T/m$ and becomes almost constant at high temperatures, for a moderately/strongly interacting Bose gas with $\alpha=0.06$/$\alpha=0.08$, $s_{\text{tot}}/T^{3}$ first increases and then decreases with $T/m$. In the high temperature regime, for a fixed $T/m$ and $y$, the entropy density decreases with increasing $\alpha$. A comparison with the entropy density in the nonrotating case from panel a shows that  $s_{\text{tot}}/T^{3}$ in a rigidly rotating gas is several orders of magnitude larger than in a nonrotating gas.}\label{fig10}
\end{figure*}
\begin{figure}[hbt]
\includegraphics[width=8cm, height=6cm]{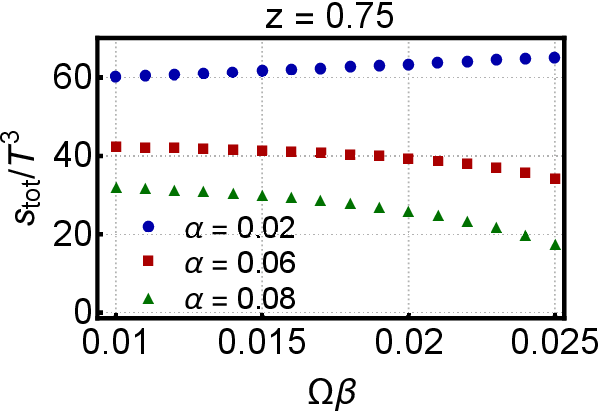}
\caption{The $\Omega\beta$ dependence of dimensionless entropy density $s_{\text{tot}}/T^{3}$ is plotted for fixed temperature $z=0.75$ and coupling $\alpha=0.02,0.06,0.08$.  Whereas for weakly interacting Bose gas, $s_{\text{tot}}/T^{3}$ slightly increases with increasing $\Omega$, for moderately and strongly interacting Bose gas, it decreases with increasing $\Omega\beta$. }\label{fig11}
\end{figure}
\begin{figure}[hbt]
\includegraphics[width=8cm, height=6cm]{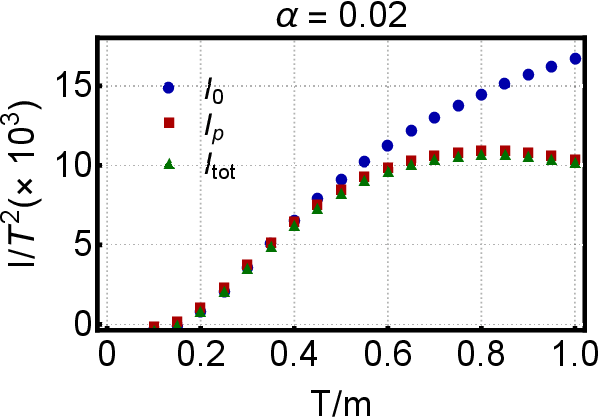}
\caption{The $T/m$ dependence of dimensionless $I_0/T^{2}, I_{\text{p}}/T^{2}$, and $I_{\text{tot}}/T^{2}$ are plotted for fixed $\alpha=0.02$. At low temperature ($T/m\lesssim 0.5$), $I_{\text{tot}}/T^{2}$ increases with increasing $T/m$. At higher temperatures $T/m>0.5$, however, $I_{\text{p}}/T^{2}$ and $I_{\text{tot}}/T^{2}$ decrease with increasing $T/m$. The $T/m$ dependence of $I_{\text{tot}}$ is mainly dominated by that of $I_{\text{p}}$.  }\label{fig12}
\end{figure}
\begin{figure*}[hbt]
\includegraphics[width=8cm, height=6cm]{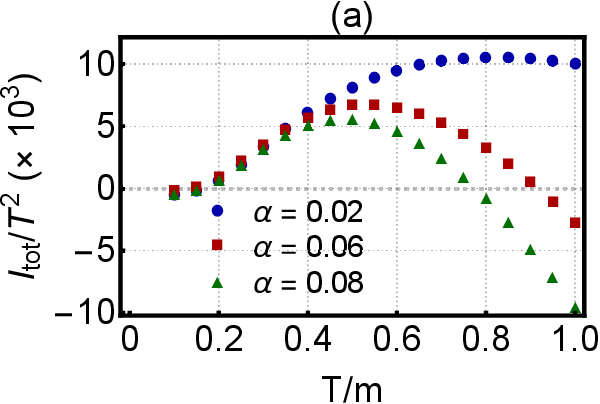}
\includegraphics[width=8cm, height=6cm]{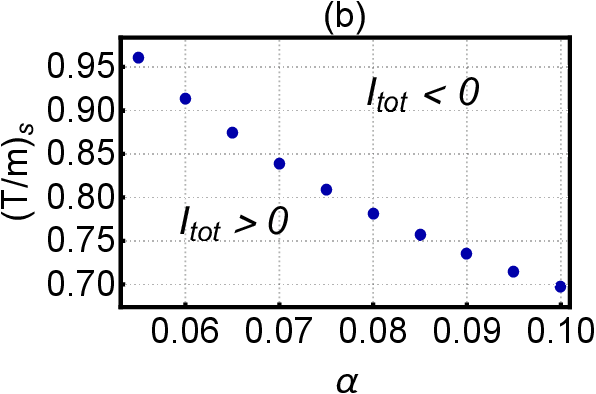}
\caption{(a) The $T/m$ dependence of dimensionless $I_{\text{tot}}/T^{2}$ is plotted for $\alpha=0.02,0.06,0.08$. Whereas for $\alpha=0.02$ (weakly interacting rotating Bose gas) $I_{\text{tot}}/T^{2}$ is always positive, for $\alpha=0.06$ and $\alpha=0.08$ (moderately and strongly interacting rotating Bose gas) $I_{\text{tot}}/T^{2}$ vanishes at certain $T/m$. These temperatures are the supervortical temperatures $(T/m)_{s}$ at which $\Omega\to \infty$. (b) The $\alpha$ dependence of the supervortical temperature $(T/m)_{s}$ is plotted. The region below (above) the blue dots corresponds to $I_{\text{tot}}<0$ ($I_{\text{tot}}>0$), and the dots indicate the supervortical temperatures for each given coupling $\alpha$.  According to these results, the supervortical temperature decreases with increasing $\alpha$, as expected from panel a.}\label{fig13}
\end{figure*}
\begin{figure*}[hbt]
\includegraphics[width=8cm, height=6cm]{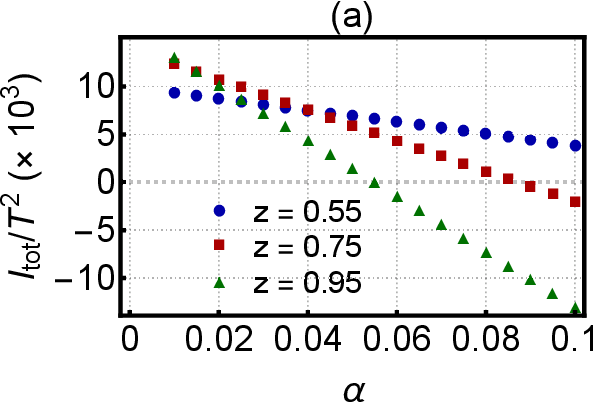}
\includegraphics[width=8cm, height=6cm]{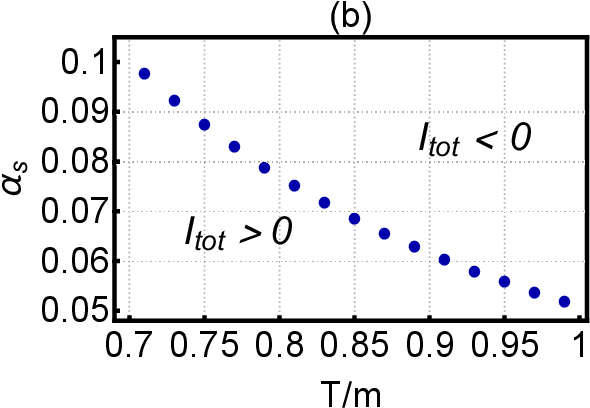}
\caption{(a) The $\alpha$ dependence of dimensionless $I_{\text{tot}}/T^{2}$ is plotted for fixed $z=0.55,0.75,0.95$. Whereas for $z=0.55$ (low temperature) $I_{\text{tot}}/T^{2}$ is positive for all values of $\alpha$, for moderate and high temperature $z=0.75$ and $z=0.95$, $I_{\text{tot}}/T^{2}$ vanishes at certain supervortical coupling, $\alpha_{s}$. (b) The $T/m$ dependence of $\alpha_{s}$ is plotted. The region below (above) the blue dots corresponds to $I_{\text{tot}}<0$ ($I_{\text{tot}}>0$), and the dots indicate the supervortical couplings for each given temperature $T/m$. According to this result, the supervortical coupling decreases with increasing temperature.}\label{fig14}
\end{figure*}
\begin{figure*}[hbt]
\includegraphics[width=8cm, height=6cm]{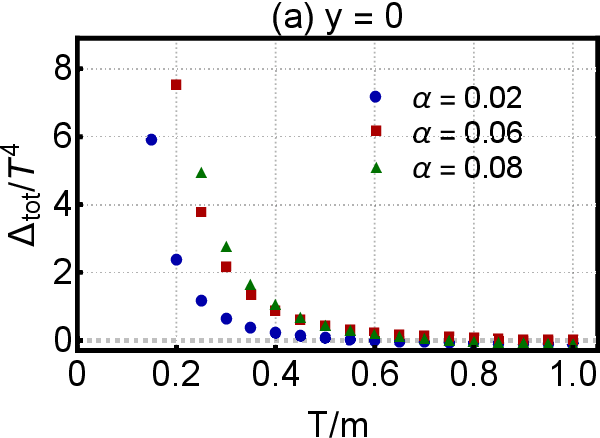}
\includegraphics[width=8cm, height=6.2cm]{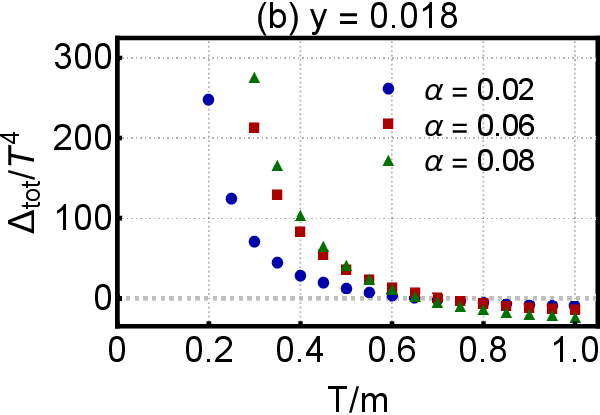}
\caption{(a) The $T/m$ dependence of dimensionless $\Delta_{\text{tot}}/T^{4}$ ($\Delta_{\text{tot}}=\epsilon_{\text{tot}}-3P_{\text{tot}}$) is plotted for a nonrotating relativistic Bose gas for fixed $\alpha=0.02,0.06,0.08$. It decreases with increasing $T/m$, and asymptotically approaches zero at high temperatures, indicating that the gas becomes ideal at high enough $T$.The slope of its fall depends on the coupling $\alpha$. (b) The $T/m$ dependence of dimensionless $\Delta_{\text{tot}}/T^{4}$ is plotted for a rotating relativistic Bose gas with $y=0.018$ for fixed $\alpha=0.02,0.06,0.08$. Although the qualitative behavior is similar to $y=0$ case in panel a, but the value of $\Delta_{\text{tot}}$ is several orders of magnitude larger than in a nonrotating Bose gas. As in the nonrotating case, $\Delta_{\text{tot}}$ decreases with increasing $T$, but it contrast to this case vanishes at certain temperature and by increasing $T$ becomes negative. This may be interpreted as a sign of thermodynamic instability of the strongly interacting medium caused by a rigid rotation.}\label{fig15}
\end{figure*}
\begin{figure}[hbt]
\includegraphics[width=8cm, height=6cm]{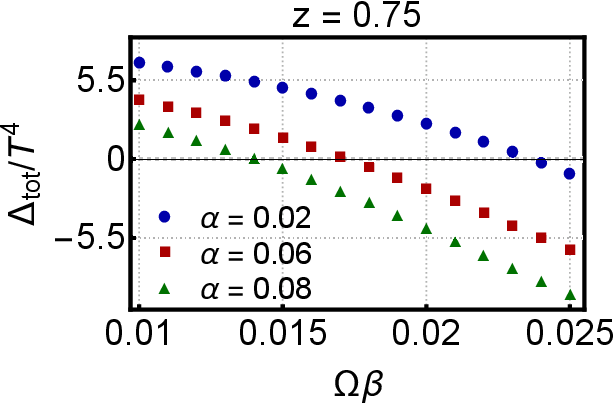}
\caption{The $\Omega\beta$ dependence of dimensionless $\Delta_{\text{tot}}/T^{4}$ ($\Delta_{\text{tot}}=\epsilon_{\text{tot}}-3P_{\text{tot}}$) is plotted for fixed $z=0.75$ and $\alpha=0.02,0.06,0.08$. It decreases with increasing $\Omega\beta$. The larger $\alpha$, the smaller the specific angular velocity is at which $\Delta_{\text{tot}}$ vanishes. Negative $\Delta_{\text{tot}}$ is a sign of thermodynamic instability of the strongly interacting medium caused by a rigid rotation.}\label{fig16}
\end{figure}
\begin{figure*}[hbt]
\includegraphics[width=8cm, height=6cm]{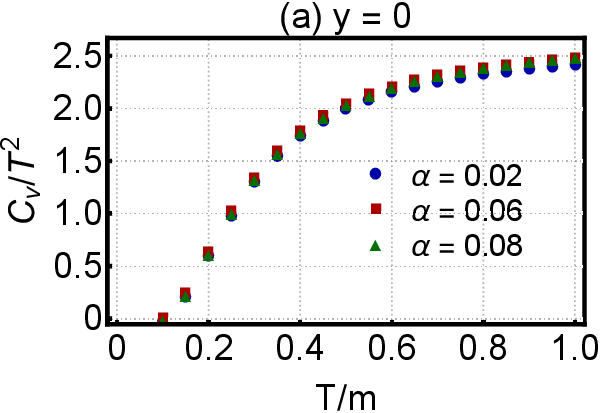}
\includegraphics[width=8cm, height=6cm]{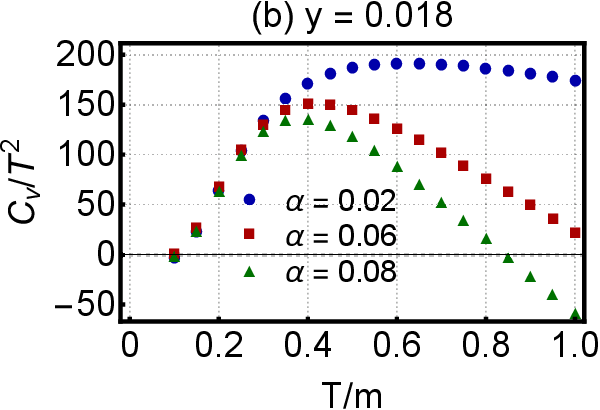}
\caption{(a) The $T/m$ dependence of dimensionless heat capacity $C_{V}/T^{2}$ for a nonrotating Bose gas is plotted for fixed $\alpha=0.02,0.06,0.08$. The heat capacity increases with increasing temperature. (b) The $T/m$ dependence of the dimensionless heat capacity $C_{V}/T^{2}$ of a rotating Bose gas is plotted for fixed $\alpha=0.02,0.06,0.08$. Comparing with the heat capacity of a nonrotating gas, $C_{V}$ in a rigidly rotating system is several orders of magnitude larger.  In contrast to the nonrotating case, $C_{V}$ first increases with increasing $T/m$ and then decreases at large temperatures. For large enough coupling $\alpha$, it vanishes at certain temperature, and becomes negative at $T$s. Negative $C_{V}$ is a sign of thermodynamic instability of the medium. It is caused by a rigid rotation in a medium with large $\alpha$.}\label{fig17}
\end{figure*}
\begin{figure}[hbt]
\includegraphics[width=8cm, height=6cm]{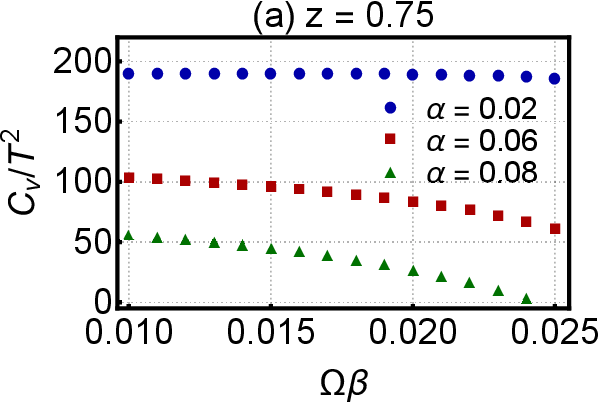}
\caption{The $\Omega\beta$ dependence of dimensionless $C_{V}/T^{2}$ is plotted for fixed temperature $z=0.75$ and $\alpha=0.02,0.06,0.08$. For a weakly interacting medium, the heat capacity is almost constant. For moderately/strongly interacting medium, however, it decreases moderately with increasing angular velocity. }\label{fig18}
\end{figure}
\begin{figure*}[hbt]
\includegraphics[width=8cm, height=6cm]{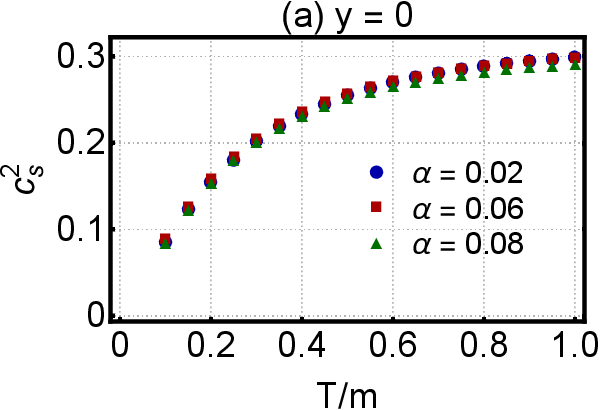}
\includegraphics[width=8cm, height=6cm]{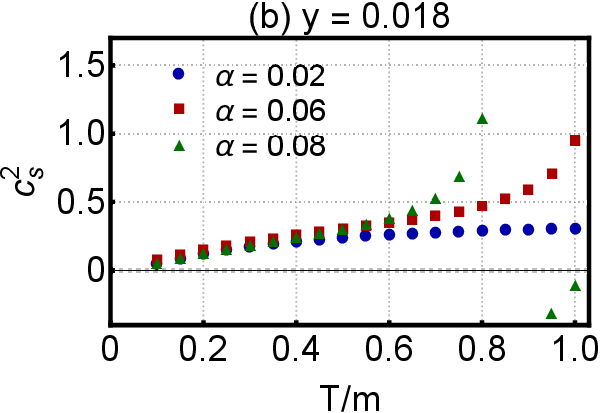}
\caption{(a) The $T/m$ dependence of the speed of sound $c_{s}^{2}$ for a nonrotating Bose gas
is plotted for fixed $\alpha=0.02,0.06,0.08$. It increases with increasing temperature and approaches asymptotically the speed of sound of an ideal gas, $c_{s}^{2}\sim 1/3$, at high temperature. (b) The $T/m$ dependence of the speed of sound $c_{s}^{2}$ for a rotating Bose gas is plotted for fixed $y=0.018$ and $\alpha=0.02,0.06,0.08$. Whereas $c_{s}$ increases with increasing $T/m$, for $\alpha=0.08$, $c_{s}^{2}$ diverges at certain temperature and at certain temperature becomes larger than the speed of light. The appearance of superluminal sound velocities ($c_{s}>1$) at high temperatures and strong $\alpha$ breaks the causality.  }\label{fig19}
\end{figure*}
\begin{figure}[hbt]
\includegraphics[width=8cm, height=6cm]{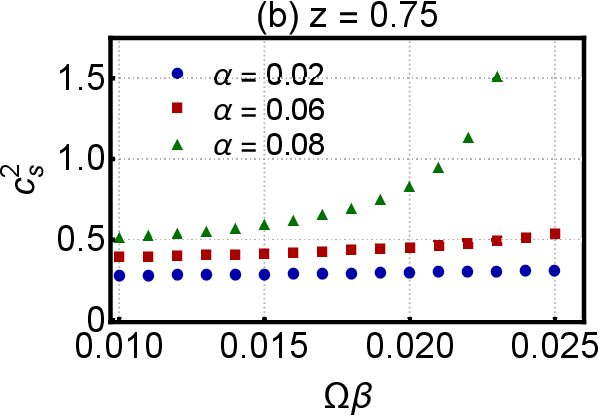}
\caption{The $\Omega\beta$ dependence of the speed of sound $c_{s}^{2}$ is plotted for fixed temperature $z=0.75$ and $\alpha=0.02,0.06,0.08$. Whereas for $\alpha=0.02$, the speed of sound is almost constant in $\Omega\beta$ and increases slightly for $\alpha=0.06$, it diverges in a strongly interacting medium with $\alpha=0.08$. The appearance of sound velocities larger than the speed of light is a sign of thermodynamic instabilities. }\label{fig20}
\end{figure}
Similarly, plugging $P_{0}$ into \eqref{M8}, the dimensionless moment of inertia $\bar{I}_{0}$ is given by
\begin{eqnarray}\label{M24}
\bar{I}_{0}(x)=-\frac{1}{64\pi^{2}}\sum_{\ell=-\infty}^{+\infty}\ell^{2}\mathcal{A}_{3,\ell}^{(2)}(x,0).
\end{eqnarray}
Plugging $P_{1}$ into \eqref{M8}, $\bar{I}_{1}$ reads
\begin{eqnarray}\label{M25}
\lefteqn{\bar{I}_{1}(x,\alpha)=-\frac{\alpha}{128\pi^{2}}\bigg\{\left(\sum_{\ell=-\infty}^{+\infty}\frac{d\mathcal{A}_{2,\ell}(x,0)}{dy}\right)^{2}
}\nonumber\\
&&+\sum_{\ell=-\infty}^{+\infty}\frac{d^{2}\mathcal{A}_{2,\ell}(x,0)}{dy^{2}}\sum_{\ell=-\infty}^{+\infty}\mathcal{A}_{2,\ell}(x,0)\bigg\},
\end{eqnarray}
with $\frac{d^{m}\mathcal{A}_{n,\ell}}{dy^{m}}$ from \eqref{M17}.
The above expressions for $\bar{I}_{0}$ and $\bar{I}_{1}$ are independent of $y$. Thus, the  summation over $\ell$ may be performed using
\begin{eqnarray*}
\sum_{\ell=-d}^{+d}\ell^n=H_{d}^{(-n)},
\end{eqnarray*}
where $H_{d}^{(n)}$ is the generalized Harmonic number \cite{wolframmath}. We need, in particular, $H_{d}^{(-1)}=d(d+1)/2$ and $H_{d}^{(-2)}=d(d+1)(2d+1)/6$.\par
Let us now consider $\bar{I}_{\text{ring}}$, which is given by plugging $P_{\text{ring}}$ into \eqref{M8}. It reads
\begin{eqnarray}\label{M26}
\bar{I}_{\text{ring}}(x,\alpha)=\frac{1}{2\pi}\sum_{\ell=-\infty}^{+\infty}\ell^{2}\bigg[\frac{\bar{\Pi}_{0}}{2x}-\left(\bar{\Pi}_{0}+x^{2}\right)^{1/2}+x\bigg], \nonumber\\
\end{eqnarray}
where again the summation over $\ell$ can be performed. To determine the zeroth and first order contribution to $\bar{C}_{V}$, we replace $P$ in \eqref{M5} with $P_{0}$ and $P_{1}$, and arrive first at
\begin{eqnarray}\label{M27}
\bar{C}_{V,0}&=&\frac{1}{16\pi^{2}}\sum_{\ell=-\infty}^{+\infty}\bigg\{12\mathcal{A}_{3,\ell}+2x^{2}\mathcal{A}_{2,\ell}+i\ell y\mathcal{A}_{3,\ell}^{(1)}\nonumber\\
&&+
4T\frac{d\mathcal{A}_{3,\ell}}{dT}+2x^{2}T\frac{d\mathcal{A}_{2,\ell}}{dT}+\frac{i\ell y}{2}T\frac{d\mathcal{A}_{3,\ell}^{(1)}}{dT}
\bigg\},\nonumber\\
\end{eqnarray}
and
\begin{eqnarray}\label{M28}
\bar{C}_{V,1}&=&-\frac{\alpha}{64\pi^{2}}\left\{\sum_{\ell=-\infty}^{+\infty}\left(\mathcal{A}_{2,\ell}+x^2\mathcal{A}_{1,\ell}+\frac{i\ell y}{4}\mathcal{A}_{2,\ell}^{(1)}\right)\right.\nonumber\\
&&\left.\times \sum_{\ell=-\infty}^{+\infty}\left(3\mathcal{A}_{2,\ell}+T\frac{d\mathcal{A}_{2,\ell}}{dT}\right)\right.\nonumber\\
&&\left.+\sum\limits_{\ell=-\infty}^{+\infty}\mathcal{A}_{2,\ell}\sum_{\ell=-\infty}^{+\infty}\left(T\frac{d\mathcal{A}_{2,\ell}}{dT}-2x^{2}\mathcal{A}_{1,\ell}\right.\right.\nonumber\\
&&\left.\left.+x^{2}T\frac{d\mathcal{A}_{1,\ell}}{dT}-\frac{i\ell y}{4}\mathcal{A}_{2,\ell}^{(1)}+\frac{i\ell y}{4}T\frac{d\mathcal{A}_{2,\ell}^{(1)}}{dT}\right)\right\}.\nonumber\\
\end{eqnarray}
Using then \eqref{M20} and
\begin{eqnarray}\label{M29}
T\frac{d\mathcal{A}_{n,\ell}^{(m)}}{dT}&=&2x^{2}\mathcal{A}_{n-1,\ell}^{(m)}+\frac{i\ell y}{2}\mathcal{A}_{n,\ell}^{(m+1)},
\end{eqnarray}
we obtain the final expression for $\bar{C}_{V,i}, i=0,1$. Finally, the ring contribution to $\bar{C}_{V}$ is given by
\begin{eqnarray}\label{M30}
\bar{C}_{V,\text{ring}}&=&-\frac{\alpha\pi}{6}\sum_{\ell=-\infty}^{+\infty}\bigg[3\bar{\zeta}_{\ell}^{1/2}-3\left(\bar{\Pi}_{0}+\bar{\zeta}_{\ell}\right)^{1/2}\nonumber\\
&&-\bar{\Pi}_{0}\left(\bar{\Pi}_{0}+\bar{\zeta}_{\ell}\right)^{-1/2}\bigg].
\end{eqnarray}
Combining the expressions corresponding to $s$ and $C_{V}$, it is possible to determine the speed of sound $c_{s}^{2}$ according to \eqref{M6}. In what follows, we focus on $T,\Omega,$ and $\alpha$ dependence of thermodynamic quantities $P,j,s,I,C_V,c_{s}^{2}$. We also compute the dimensionless energy density $\bar{\epsilon}$ using $\bar{\epsilon}=-\bar{P}+\bar{s}$ and explore the $T,\Omega,\alpha$ dependence of the interaction measure $\bar{\Delta}\equiv\bar{\epsilon}-3\bar{P}$ \cite{endrodi-book}. We use  following notations: $y=\Omega\beta, z\equiv T/m$, $P_{\text{p}}\equiv P_{0}+P_{1}$, and $\bar{\mathcal{X}}_{\text{tot}}\equiv \bar{\mathcal{X}}_{0}+\bar{\mathcal{X}}_{1}+\bar{\mathcal{X}}_{\text{ring}}$, with $\mathcal{X}=\{P,s,j,I,\epsilon\}$.
\subsubsection{Results}\label{sec4B2}
In Sec. \ref{sec3A}, the exact and HTE expressions of $V_{\text{eff}}^{(0)T}$ are given in \eqref{A13} and \eqref{A15}. According to \eqref{M1}, the exact and HTE expressions of the zeroth order pressure $P_{0}$ are determined from these expressions. In Fig. \ref{fig5}, we compare the $T/m$ dependence of the exact (blue dots) and the HTE (red squares) of the dimensionless $P_{0}/T^{4}$ for a nonrotating $y=0$ [Fig. \ref{fig5}(a)] and rotating [Fig. \ref{fig5}(b)] relativistic Bose gas with $y=0.018$. As it is demonstrated, in both cases two expressions coincide for $z\geq 0.4$. This fixes the reliability regime for HTE. A comparison between two plots shows that the rotation increases the pressure $P_{0}$ up to several orders of magnitude.
\par
In Fig. \ref{fig6}, the $T/m$ and $\Omega\beta$ dependence of dimensionless $P_{\text{ring}}/T^{4}$ is presented for $\alpha=0.02,0.06,0.08$ and fixed $y=0.018$ [Fig. \ref{fig6}(a)] as well as $z=0.75$ [Fig. \ref{fig6}(b)]. It is shown that  $P_{\text{ring}}/T^{4}$ increases with increasing $T/m$ and $\Omega\beta$. The $\alpha$ dependence of $P_{\text{ring}}/T^{4}$ is, however, nontrivial. It is plotted in Fig. \ref{fig7} for fixed $y=0.018$ and various temperatures $z=0.7,0.8,0.9$. It turns out that at low temperature ($z=0.7$) $P_{\text{ring}}/T^{4}$ is negative and decreases with increasing $\alpha$, while at high temperatures ($z=0.8,0.9$), it is first negative, then increases with $\alpha$ and becomes positive for $\alpha\sim 1$.
\par
In Fig. \eqref{fig8}(a), the $T/m$ dependence of $P_{\text{p}}/T^4$ (blue circles) and $P_{\text{tot}}/T^4$ (red squares) are plotted for fixed $y=0.018$ and $\alpha=0.02$.
At $T/m>0.4$, $P_{\text{p}}$ and $P_{\text{tot}}$ are both positive and their difference ($P_{\text{tot}}-P_{\text{p}}=P_{\text{ring}}$) becomes negligible. The same is also true for their $\Omega\beta$ dependence once $z$ and $\alpha$ are fixed [see Fig. \ref{fig8}(b)]. According to this plot, $P_{\text{p}}$ and $P_{\text{tot}}$ increases with increasing $\Omega\beta$, and their difference decreases with increasing $\Omega\beta$. These results are in complete agreement with the results from Fig. \eqref{fig6}, where the absolute value of $P_{\text{ring}}$ decreases with increasing temperature and angular velocity.
\par
Let us now consider the $T/m$ and $\Omega\beta$ dependence of the dimensionless angular momentum density $\bar{j}_{\text{tot}}$, which includes the contribution from $\bar{j}_{0},\bar{j}_{1}$, and $\bar{j}_{\text{ring}}$. In Fig. \ref{fig9}(a), the $T/m$ dependence of $\bar{j}_{tot}$ is plotted for fixed $y=0.018$ and three different coupling $\alpha=0.02,0.06,0.08$. According to this plot, $\bar{j}_{tot}$ first increases with $T/m$, at some temperature becomes maximum, and then decreases with increasing temperature. The position of the maxima and the point at which $\bar{j}_{tot}$ vanishes and then changes its sign depends on the strength of the coupling $\alpha$. The larger $\alpha$, the smaller is the maximum of $\bar{j}_{tot}$. This specific $T$ dependence of $\bar{j}_{tot}$ may be interpreted as a sign of thermodynamic instability in the medium, which turns out to be more probable for strong couplings $\alpha$.
\\
In Fig. \eqref{fig9}(b), we explore the $\Omega\beta$ dependence of $\bar{j}_{tot}$ for fixed temperature $z=0.75$ and $\alpha=0.02,0.06,0.08$. It turns out that for a weakly interacting relativistic Bose gas, $\bar{j}_{tot}$ increases with $\Omega\beta$, whereas for a moderately interacting one, it is first positive and remains almost constant for $\Omega\beta\sim 0.018$. Afterward it becomes zero and changes its sign for larger $\Omega\beta$. For $\alpha=0.08$ and $z=0.75$, however, it turns out to be negative, in accordance with the plots from Fig. \ref{fig9}(a). The plot from Fig. \ref{fig9}(b) indicates that $\bar{j}_{tot}$ is not linear in $\Omega\beta$, especially for larger $\Omega\beta$s.
\par
In Fig. \eqref{fig10}, the $T/m$ dependence of the dimensionless entropy density $\bar{s}_{\text{tot}}$ for a nonrotating ($y=0$) and a rotating ($y=0.018$) relativistic Bose gas is compared. As in the previous case, $\bar{s}_{\text{tot}}$ receives contributions from $\bar{s}_{0}$, $\bar{s}_{1}$, and $\bar{s}_{\text{ring}}$. As it is shown in Fig. \ref{fig10}(a), in a nonrotating gas, the entropy increases with increasing temperature and changing the strength of the interaction $\alpha$ has practically no effect on this behavior. For a rotating gas, however, the coupling $\alpha$ substantially affects the $T$ dependence of $\bar{s}_{\text{tot}}$. Apart from the fact that its value increases up to several orders of magnitude due to rotation, the $T$ dependence of $\bar{s}_{\text{tot}}$ in a weakly interacting medium is similar to the $T$ dependence of a nonrotating gas. In contrast, for larger values of $\alpha$, $\bar{s}_{\text{tot}}$
increases first with increasing temperature, exhibits a maximum at a certain temperature decreases with increasing $T$. Hence, an interplay between the coupling $\alpha$ and angular velocity $\Omega$ in the final expression for $\bar{s}_{\text{tot}}$ leads to a more ordered system at high temperature. In what follows, we show that this counterintuitive $T$ dependence of the entropy density leads to two novel phenomena in a rigidly rotating relativistic Bose gas: (i) the emergence of negative heat capacity and (ii) the appearance of superluminal sound velocities at high enough temperatures and large enough couplings. Both effects are signs of thermodynamic instability.
\par
In Fig. \ref{fig11}, the $\Omega\beta$ dependence of the dimensionless entropy density is plotted for a fixed temperature $z=0.75$ and various couplings $\alpha=0.02,0.06,0.08$. In the weakly interacting case ($\alpha=0.02$), $s_{\text{tot}}/T^{3}$ increases slightly with increasing $\Omega\beta$. In contrast, for a moderately/strongly interacting medium, $s_{\text{tot}}/T^{3}$ decreases with $\Omega\beta$. According to these results, we conclude that the coupling constant $\alpha$ plays an important role on the $T/m$ as well as $\Omega\beta$ dependence of the total entropy density.  Moreover, in general, for fixed temperature and angular velocity, the total entropy density decreases with increasing coupling $\alpha$.
\par
Let us now consider the temperature dependence of the dimensionless moment of inertia $I_{\text{tot}}/T^{2}$. In Fig. \ref{fig12}, the $T/m$ dependence of $\bar{I}_{0}$ (blue dots), $\bar{I}_{p}$ (red squares), and $\bar{I}_{\text{tot}}=I_{0}+I_{p}$ (green triangles) is plotted. For a weakly interacting relativistic Bose gas with $\alpha=0.02$, $\bar{I}_{\text{tot}}$ increases with increasing temperature. In addition, its $T/m$ dependence turns out to be mainly dominated by that of $\bar{I}_{\text{p}}=\bar{I}_{0}+\bar{I}_{1}$.
\\
The fact that $\bar{I}_{\text{tot}}$ is positive for the whole interval $T/m\in[0,1]$ is only true for a weakly interacting gas. In Fig. \ref{fig13}(a), we explore the $T/m$ dependence of the dimensionless $\bar{I}_{\text{tot}}$ for a weakly, moderately, and strongly interacting rotating Bose gas with couplings $\alpha=0.02,0.06$, and $\alpha=0.08$, respectively. It turns out that in a weakly interacting medium $\bar{I}_{\text{tot}}$ is positive in the whole interval of temperature, while in a moderately/strongly interacting gas, it first increases with $T/m$, has then a maximum at some moderate temperature, and eventually falls and changes sign at high temperatures.
Following the terminology introduced recently in \cite{chernodub2023-1,chernodub2023-2}, we refer to temperatures at which $\bar{I}_{\text{tot}}$ vanishes as `'supervortical temperatures'' $z_{s}\equiv (T/m)_{s}$. According to the results in Fig. \ref{fig13}(a), at $z<z_{s}$ ($z>z_{s}$) the total moment of inertia is positive (negative). In Fig. \ref{fig13}(b), the $\alpha$ dependence of supervortical temperatures $(T/m)_{s}$ is plotted. The blue dots indicate supervortical temperatures for each given $\alpha$. The region below (above) the blue dots corresponds to $\bar{I}_{\text{tot}}>0$ ($\bar{I}_{\text{tot}}<0$).
\par
We also examine the $\alpha$ dependence of $\bar{I}_{\text{tot}}/T^{2}$ for fixed temperatures $z=0.55,0.75,0.95$ [see Fig. \ref{fig14}(a)]. Whereas for $z=0.55$, the dimensionless moment of inertia is positive for $0<\alpha<0.1$, at higher temperatures, there exists a certain supervortical coupling for which $\bar{I}_{\text{tot}}$ vanishes. Consequently, for $\alpha<\alpha_{s}$ ($\alpha>\alpha_{s}$)  $\bar{I}_{\text{tot}}$ turns out to be positive (negative). In Fig. \ref{fig14}(b), the $T/m$ dependence of $\alpha_{s}$ is plotted. It decreases with increasing temperatures as expected from Fig. \ref{fig14}(a). Again, the blue dots indicate the supervortical couplings for each given temperature $T/m$ and the region below (above) the blue dots corresponds to $\bar{I}_{\text{tot}}>0$ ($\bar{I}_{\text{tot}}<0$).
\par
At this stage a couple of remarks concerning $I=0$ and $I<0$ are in order. Using $j=I\Omega$ from \eqref{M9} and assuming that $j=\text{const.}$, a vanishing moment of inertia leads to an extremely large angular velocity $\Omega$. This is why the term `'supervortical" is used in \cite{chernodub2023-1,chernodub2023-2}. A negative moment of inertia, however, means that by applying an external angular momentum $\boldsymbol{J}$, the system rotates with an angular velocity $\boldsymbol{\Omega}$ directed antiparallel to $\boldsymbol{J}$ (see Fig. \ref{fig1} for a visualization of this situation). Thus, a negative moment of inertia may indicate a thermodynamic instability in a rigidly rotating medium \cite{chernodub2023-1,chernodub2023-2}.
\par
In Fig. \ref{fig15}, we explore the $T$ dependence of dimensionless $\bar{\Delta}= \bar{\epsilon}-3\bar{P}$ for a nonrotating $y=0$ and a rotating $y=0.018$ relativistic Bose gas. This quantity is a measure
for the ideality of a relativistic medium, as $\epsilon=3P$ is the equation of state of an ideal (Bose) gas. As it is  demonstrated in Fig. \ref{fig15}(a), $\bar{\Delta}_{\text{tot}}$ decreases with increasing temperature. The slope of its fall depends slightly on the coupling $\alpha$. This result indicates that at high enough temperature the nonrotating Bose gas behaves as an ideal gas, as $\bar{\Delta}_{\text{tot}}$ approaches asymptotically to zero. For a rotating medium, apart from the fact that $\bar{\Delta}_{\text{tot}}$ is up to several orders of magnitude larger than in a nonrotating medium, it decreases with increasing temperature [see Fig. \ref{fig15}(b)]. However, in contrast to the nonrotating case, it vanishes at certain temperature, and becomes negative afterward. This is a sign of a thermodynamic instability caused by a rigid rotation in a strongly interacting relativistic Bose gas.
\par
The $\Omega\beta$ dependence of $\bar{\Delta}_{\text{tot}}$ is plotted in Fig. \ref{fig16} for fixed temperature $z=0.75$ and $\alpha=0.02,0.06,0.08$. According to this plot, $\bar{\Delta}_{\text{tot}}$ decreases with increasing $\Omega\beta$. At some specific $\Omega\beta$, it vanishes and becomes negative. The stronger the coupling constant, the lower the angular velocity is at which $\bar{\Delta}_{\text{tot}}$ vanishes and the system becomes unstable.
\par
In Fig. \ref{fig17}, the temperature dependence of dimensionless heat capacity $C_{V}/T^2$ is plotted for a nonrotating $y=0$ and rotating ($y=0.018$) relativistic Bose gas. We used \eqref{M5} to determine $C_{V}$. According to the plot in Fig. \ref{fig17}(a), in the nonrotating medium, the heat capacity increases with increasing temperature. In a rotating medium, however, the $T/m$ dependence of $\bar{C}_{V}$ depends significantly on $\alpha$ [see Fig. \ref{fig17}(b)]. Whereas for $\alpha=0.02$ (weakly interacting medium), the heat capacity is always positive and its $T$ dependence is more or less similar to the case of a nonrotating gas, for a moderately and strongly interacting gas with $\alpha=0.06$ and $\alpha=0.08$, $\bar{C}_{V}$ decreases with increasing temperature. For $\alpha=0.08$ at $T/m\sim 0.82$, it vanishes and then becomes negative at  $T/m>0.82$. Let us notice that when a system possesses a negative heat capacity, its temperature decreases by supplying heat. Same counterintuitive behavior appears in a rigidly rotating Bose gas, and is affected by the strength of the interaction in the medium.
\\
The $\Omega\beta$ dependence of the dimensionless heat capacity is plotted in Fig. \ref{fig18} for fixed temperature $z=0.75$ and $\alpha=0.02,0.06,0.08$. In the weakly interacting case $\alpha=0.02$, $\bar{C}_{V}$ is positive and almost constant in $\Omega\beta$. For $\alpha=0.06$ and $\alpha=0.08$, however, it slightly decreases with increasing $\Omega\beta$. For large coupling $\alpha=0.08$, it vanishes at some large $\Omega\beta$. For fixed $T$ and $\Omega$, $\bar{C}_{V}$ decreases with increasing $\alpha$.
\par
Using the data corresponding to the entropy density and heat capacity, it is possible to determine the sound velocity $c_{s}$ according to \eqref{M6}. In Fig. \eqref{fig19}, the $T/m$ dependence of $c_{s}^{2}$ is plotted for fixed $\alpha=0.02,0.06,0.08$ as well as $y=0$ [Fig. \ref{fig19}(a)] and $y=0.018$ [Fig. \ref{fig19}(b)]. According to these results, the speed of sound of a nonrotating Bose gas increases with increasing $T$ and approaches asymptotically the speed of sound of a free relativistic gas,  $c_{s}^{2}=1/3$. In the absence of rotation, different choices of $\alpha$ does not affect this behavior too much. In a rotating medium, however, the situation is different. Whereas,
according to the results in Fig. \ref{fig19}(b), the temperature dependence of $c_{s}^{2}$ is more of less similar to the nonrotating case, for a moderately interacting gas with $\alpha=0.06$, $c_{s}^{2}$ increases with $T$ but it passes $1/3$ at high temperature and becomes almost equal to the speed of light at $T/m\sim 1$. For strong coupling $\alpha=0.08$, the sound velocity increases very fast, so that at $T/m=0.8$ is given by $c_{s}=1.07>1$. This breaks the causality and is an indication that a strongly interacting rotating Bose gas becomes unstable at high temperature.
\\
The $\Omega\beta$ dependence of $c_{s}^{2}$ is explored in Fig. \ref{fig20} for fixed temperature $z=0.75$ and $\alpha=0.02,0.06,0.08$. For a weakly interacting Bose gas, the speed of sound is lower than $c_{s}^{2}=1/3$. It increases slightly with increasing $\Omega\beta$, but never becomes larger than the speed of light. The same is also true for a moderately interacting medium with $\alpha=0.06$. For $\alpha=0.08$, however, $c_{s}^{2}$ increases very fast with increasing $\Omega\beta$ and reaches $c_{s}\sim 1$ at $\Omega\beta\sim 0.022$. Afterward the system becomes unstable because of broken causality for larger values of angular velocity. Let us notice that $\alpha=0.08$ corresponds to $\lambda\sim 0.78<1$, which is still reliable for a perturbative expansion. The above results show that such a strongly interacting Bose gas become unstable either at large temperatures or large angular velocities once the system is subjected to a rigid rotation.
\section{Concluding remarks}\label{sec5}
We studied the effect of a rigid rotation on the thermodynamic properties of a relativistic Bose gas. First, we determined the perturbative thermodynamic potential up to one-loop order, which together with the nonperturbative ring potential was used to compute the thermodynamic quantities in this approximation. To do this, we considered the Lagrangian density of a CKG model in the presence of a rigid rotation. We utilized the solution of the corresponding equation of motion to derive the free propagator of this model using the Fock-Schwinger method. The free propagator allowed us to determine the thermodynamic potential of this model, including zeroth and one-loop perturbative contributions as well as nonperturbative ring potential. We presented analytical expressions for these quantities and showed, in particular, that the angular velocity $\Omega$ plays effectively the role of a chemical potential, as anticipated from literature. Additionally, we performed an appropriate high temperature expansion and presented the corresponding results to the total thermodynamic potential in this approximation. This potential was then used to determine several thermodynamic quantities, including the pressure, entropy density, angular momentum density, heat capacity, speed of sound, and the moment of inertia of this rotating relativistic Bose gas. We numerically explored the $T$ and $\Omega$ dependence of these quantities.
\par
By comparing  the exact expression of $P_{0}$, arising from the zeroth order thermodynamic potential, with the high temperature expanded expressions corresponding to it, we determined the high temperature regime of this model to be $T/m\geq 0.4$. We showed that $P_{0}$ of a rotating relativistic Bose gas is much higher than $P_{0}$ for a nonrotating gas.  We then focused on the one-loop and ring contributions to the total pressure $P_{\text{tot}}$. As the ring potential is negative in the whole interval of $T$ and $\Omega$, and as it increases with increasing $T$ and $\Omega$, its effect reduces in high temperature and frequency regimes. Hence, in this regime, the $(T,\Omega)$ dependence of the total pressure is mainly dominated by the $(T,\Omega)$ dependence of $P_{1}$, including the zeroth and one-loop contributions to $P_{\text{tot}}$.
Apart from $(T,\Omega)$ dependence of the pressure, we focused on its $\alpha=\lambda/\pi^{2}$ dependence. Here, $\lambda$ is the coupling constant of the model, which appears in the corresponding Lagrangian density. We showed that the ring pressure exhibits a nonlinear dependence on $\alpha$.
\par
Regarding the  $(T,\Omega)$ dependence of the angular momentum and entropy densities, $j_{\text{tot}}$ and $s_{\text{tot}}$, for fixed $(\Omega\beta, T/m)$ and $\alpha$, we distinguish three different types of behavior in three different regimes of $\alpha$. Whereas in the weakly interacting regime $0<\alpha\leq 0.05$, $j_{\text{tot}}$ is positive and $s_{\text{tot}}$ increases with increasing temperature and angular velocity, in the moderately and strongly interacting regimes $\alpha\in[0.05,0.07]$ and $\alpha\in[0.07,0.1]$, $j_{\text{tot}}$ becomes negative, in particular, in the high temperature regime and $s_{\text{tot}}$ decreases with increasing temperature. This is an effect mainly caused by the rigid rotation, as, for instance, the entropy density of a nonrotating relativistic Bose gas increases with increasing $T$, as expected.
\par
Being directly related to $j_{\text{tot}}$ through its definition in \eqref{M9}, the $T$ dependence of the moment of inertia $I_{\text{tot}}$ is also affected by $\alpha$. Whereas in the weakly interacting regime, it is positive, it becomes negative in a moderately and strongly interacting medium after certain temperature. The specific temperature at which  $I_{\text{tot}}$ vanishes, was referred to as the supervortical temperature, $T_{s}$. We demonstrated in Fig. \ref{fig12}(b) that $T_{s}$ decreases with increasing $\alpha$. Apart from $T_{s}$, we defined a supervortical coupling $\alpha_{s}$, and showed in Fig. \ref{fig13}(b) that $\alpha_{s}$ decreases with increasing temperature. Interpreting $\boldsymbol{j}=I\boldsymbol{\Omega}$ as the linear response to $\Omega$, the moment of inertia $I$ plays the role of the susceptibility of the medium corresponding to rotation.  As it is demonstrated in Fig. \ref{fig1}, $I>0$ ($I<0$) means that by applying an angular momentum $\boldsymbol{j}$, the system rotates with $\boldsymbol{\Omega}$ parallel (antiparallel) to $\boldsymbol{j}$ and a vanishing moment of inertia leads to $\Omega\to \infty$ once $j$ is assumed to be finite. Similar counterintuitive effect is also observed in the temperature dependence of the heat capacity $C_{V}$. Whereas $C_{V}$ is positive in a weakly interacting Bose gas under rotation, in a moderately interacting gas, it decreases with increasing temperature, and in a strongly interacting Bose gas, it vanishes at some finite temperature and becomes negative with increasing temperature. Negative $C_{V}$ means that although a system receives heat, but its temperature decreases. Its occurrence is a sign of thermodynamic instability in a medium. Here, this instability is caused by rigid rotation.
\par
Another noticeable effect that occurs once the relativistic Bose gas is strongly interacting and rigidly rotates, is the appearance of superluminal sound velocities at high temperatures and for large angular velocities (see Figs. \ref{fig19} and \ref{fig20}). According to \eqref{M6}, the sound velocity $c_{s}$ is defined in terms of the entropy density and heat capacity. Its $(T,\Omega,\alpha)$ dependence is thus directly related to the $(T,\Omega,\alpha)$ dependence of the entropy density. It thus seems that a relativistic Bose gas under rigid rotation becomes thermodynamically unstable in the strong coupling regime $\alpha\in [0.07,0.1]$, in which (because of $\lambda<1$) perturbative computation is still possible. We thus conclude that the above-mentioned instabilities are caused by an interplay between free parameters $(T,\Omega,\alpha)$.
\par
In summary, the analysis of the thermodynamic properties of the rigidly rotating relativistic Bose gas revealed interesting behavior at high temperatures and large coupling constants. The appearance of thermodynamic instabilities, such as zero and negative values of the moment of inertia and heat capacity, suggested the presence of unique phenomena such as supervorticity and provided insight into the complex behavior of a rigidly rotating system. It would be interesting to extend this work to a relativistic Fermi gas and eventually generalize it to the QGP produced in relativistic HICs. First attempt in this direction is made in \cite{mameda2023}. In general, the study of such systems not only enriches our knowledge of fundamental physics, but may offer potential applications in diverse fields such as condensed matter physics \cite{negativeinertiaconverter} and astrophysics \cite{mielke1998,schunck2003, gravwaveneginertia}.
\begin{appendix}
\section{Free bosonic propagator in momentum space}\label{appA}
The free boson propagator in the coordinate space is given by \eqref{S20}. The corresponding propagator in the Fourier space is determined by
\begin{eqnarray}\label{appA1}
D_{\ell\ell'}^{(0)}(p,p')=\int d^{4}xd^{4}x'D_{0}(x,x')\phi_{\ell}(x,p)\phi_{\ell'}(x',p'), \nonumber\\
\end{eqnarray}
with $d^{4}x=dtd\varphi dzrdr$ in the cylindrical coordinate system. Plugging $D_{0}(x,x')$ from \eqref{S21} and $\phi_{\ell}(x,p)$ from \eqref{S10} into \eqref{appA1}, we arrive first at
\begin{widetext}
\begin{eqnarray}\label{appA2}
D_{\ell\ell'}^{(0)}\left( p,{p}' \right)&=&\sum\limits_{n=-\infty}^{+\infty}{\int{dtd{t}'d\varphi d{\varphi }'dzd{z}'rdr{r}'d{r}'}}\nonumber\\
		 &&\times \int{\frac{dEd{{k}_{z}}d{{k}_{\perp }}{{k}_{\perp }}}{{{\left( 2\pi  \right)}^{3}}}\frac{{{e}^{-iE\left( t-{t}' \right)+in \Omega \left( t-{t}' \right)+i{{k}_{z}}\left( z-{z}' \right)+in \left( \varphi -{\varphi }' \right)}}}{{{E}^{2}}-k_{\perp }^{2}-k_{z}^{2}-{{m}^{2}}+i\epsilon}{{J}_{n }}\left( {{k}_{\perp }}r \right){{J}_{n }}\left( {{k}_{\perp }}{r}' \right)}\nonumber \\
		 &&\times {{e}^{+i{{p}_{0}}t-i{\ell }\varphi -i{{p}_{z}}z}}{{J}_{{{\ell }}}}\left( {{p}_{\perp }}r \right)\times e^{-ip'_{0}t+i\ell '\varphi'+ip'_{z}z'}J_{\ell'}\left(p'_{\perp }r' \right).
\end{eqnarray}
\end{widetext}
To perform the integrations over $t$ and $z$, we use
\begin{eqnarray}\label{appA3}
\int dt e^{-i\left(E-\left(n\Omega+p_{0}\right)\right)t}&=&2\pi\delta\left(E-\left(n\Omega+p_{0}\right)\right), \nonumber\\
\int dze^{i\left(k_{z}-p_{z}\right)z}&=&2\pi\delta\left(k_{z}-p_{z}\right).
\end{eqnarray}
Integration over $t'$ and $z'$ are performed similarly. The integral over $\varphi$ yields
\begin{eqnarray}\label{appA4}
\int_{0}^{2\pi}d\varphi e^{i\left(n-\ell\right)\varphi}=2\pi\delta_{n\ell}.
\end{eqnarray}
Similarly, the integration over $\varphi'$ leads to
\begin{eqnarray}\label{appA5}
\int_{0}^{2\pi}d\varphi e^{i\left(n-\ell'\right)\varphi}=2\pi\delta_{n\ell'}.
\end{eqnarray}
Because of the summation over $n$ in \eqref{appA2}, \eqref{appA4} and \eqref{appA5} result in  $\ell=\ell'=n$. It is thus possible to perform the integration over $r$ and $r'$ by making use of
\begin{eqnarray}\label{appA6}
\int_{0}^{\infty }drrJ_{\ell }\left(k_{\perp }r \right)J_{\ell }\left(p_{\perp }r \right)&=&\frac{1}{k_{\perp }}\delta \left(k_{\perp }-p_{\perp } \right),\nonumber\\
\int_{0}^{\infty }dr'r'J_{\ell }\left(k_{\perp }r' \right)J_{\ell }\left(p'_{\perp }r' \right)&=&\frac{1}{k_{\perp }}\delta \left(k_{\perp }-p'_{\perp } \right).\nonumber\\
\end{eqnarray}
Plugging \eqref{appA3}, \eqref{appA4}, \eqref{appA5} and \eqref{appA6} into \eqref{appA2}, and performing the integration over $E, k_{z},k_{\perp}$ and the summation over $n$, we arrive at
\begin{eqnarray}\label{appA7}
D_{\ell\ell'}^{(0)}(p,p')=\frac{(2\pi)^{3}\widehat{\delta}_{\ell,\ell'}(p_{0},p_{z},p_{\perp};p^{\prime}_{0},p^{\prime}_{z},p^{\prime}_{\perp})}{\left(p_{0}+\ell\Omega\right)^{2}-p_{\perp}^{2}-p_{z}^{2}-m^{2}+i\epsilon},\nonumber\\
\end{eqnarray}
with
\begin{eqnarray}\label{appA8}
\widehat{\delta}_{\ell,\ell'}(p_{0},p_{z},p_{\perp};p^{\prime}_{0},p^{\prime}_{z},p^{\prime}_{\perp})&=&\frac{1}{p_{\perp}}\delta\left(p_{0}-p^{\prime}_{0}\right)\delta\left(p_{z}-p^{\prime}_{z}\right)\nonumber\\
&&\times
\delta\left(p_{\perp}-p^{\prime}_{\perp}\right)\delta_{\ell\ell'},
\end{eqnarray}
[see \eqref{S22} and \eqref{S23}].
\section{High temperature expansion of $\bs{V_{\text{eff}}^{(0)T}}$}\label{appB}
\setcounter{equation}{0}
In this appendix, we derive \eqref{A15} which arises by an appropriate HTE of the $T$ dependent part of the zeroth order correction to the thermodynamic (effective) potential \eqref{A7}. To this purpose, we use the method introduced in \cite{toms-book, laine-book}, where the thermodynamic potential of a free relativistic Bose gas with a finite chemical potential $\mu$ is expanded in the orders of $m\beta$, with $\beta=T^{-1}$ and $\mu<m$.\footnote{Here, $\mu$ is assumed to be positive.}
\par
Let us consider $V_{\text{eff}}^{(0)T}$ from \eqref{A7},
\begin{eqnarray}\label{appB1}
V_{\text{eff}}^{(0)T}&=&T\sum_{\ell=-\infty}^{+\infty}\int \frac{dp_{z}p_{\perp}dp_{\perp}}{(2\pi)^{2}}\nonumber\\
&&\times \bigg[\ln\left(1-e^{-\beta \left( \omega +\ell \Omega  \right)} \right)+\ln \left( 1-e^{-\beta\left( \omega -\ell \Omega  \right)} \right)\bigg],\nonumber\\
\end{eqnarray}
and separate the summation over $\ell$ into the contribution from $\ell=0$ and $\ell\neq 0$ to $V_{\text{eff}}^{(0)T}$. The resulting expression is then given by
\begin{eqnarray}\label{appB2}
V_{\text{eff}}^{(0)T}=2\left(\mathcal{I}_{1}+\mathcal{I}_{2}\right),
\end{eqnarray}
with
\begin{eqnarray}\label{appB3}
\mathcal{I}_{1}&\equiv&\int\frac{dp_{z}p_{\perp}dp_{\perp }}{\left( 2\pi  \right)^{2}}\ln \left( 1-
e^{-\beta \omega } \right),\nonumber\\
{{\mathcal{I}}_{2}}&\equiv&\sum\limits_{\ell =1}^{+\infty }\int\frac{dp_{z}p_{\perp}dp_{\perp }}{\left( 2\pi  \right)^{2}}\nonumber\\
&&\times\left[ \ln \left( 1-e^{-\beta \left( \omega +\ell \Omega  \right)} \right)+\ln \left( 1-e^{-\beta \left( \omega -\ell \Omega  \right)} \right) \right].\nonumber\\
\end{eqnarray}
Here, $\omega^{2}=p_{\perp}^{2}+p_{z}^{2}+m^{2}$. To evaluate the $\Omega$-independent part of $V_{\text{eff}}^{(0)T}$, $\mathcal{I}_{1}$, we use
\begin{eqnarray}\label{appB4}
\ln (1-x)=-\sum\limits_{k=1}^{+\infty }\frac{x^{k}}{k},
\end{eqnarray}
and arrive first at
\begin{eqnarray}\label{appB5}
\mathcal{I}_{1}=-\sum\limits_{k=1}^{+\infty }\frac{1}{k}\int\frac{dp_{z}p_{\perp}dp_{\perp }}{\left( 2\pi  \right)^{2}}e^{-\beta \omega k}.
\end{eqnarray}
Using, at this stage, the Mellin transformation of the exponential function in \eqref{appB5}, we obtain
\begin{eqnarray}\label{appB6}
e^{-\beta\omega k}=\frac{1}{2\pi i}\int_{c-i\infty }^{c+i\infty }dz\,\Gamma \left( z \right)
\left(\beta k\right)^{-z}\left(\omega^{2}\right)^{-z/2}.\nonumber\\
\end{eqnarray}
Plugging then
\begin{eqnarray}\label{appB7}
\left(\omega^{2}\right)^{-z/2}=\frac{1}{\Gamma(z/2)}\int_{0}^{\infty}dt t^{\frac{z}{2}-1}e^{-\omega^{2}t},
\end{eqnarray}
into \eqref{appB6} and the resulting expression into \eqref{appB5}, we arrive at
\begin{eqnarray}\label{appB8}
\mathcal{I}_{1}&=&-\frac{1}{2\pi i}\int_{c-i\infty}^{c+i\infty}dz\zeta(z+1)\frac{\Gamma(z)}{\Gamma\left(z/2\right)}\beta^{-z}\nonumber\\
&&\times \int_{0}^{\infty}dt t^{\frac{z}{2}-1}e^{-m^{2}t}\int\frac{dp_{z}p_{\perp}dp_{\perp}}{\left(2\pi\right)^{2}}e^{-\left(p_{z}^{2} +p_{\perp}^{2}\right)t},\nonumber\\
\end{eqnarray}
where $\zeta(z)$ is the Riemann $\zeta$-function. It arises from
\begin{equation}\label{appB9}
\sum\limits_{k=1}^{+\infty }{{{k}^{-\left( 1+z \right)}}}=\zeta \left( 1+z \right),
\end{equation}
that is used to perform the summation over $k$ in \eqref{appB8}. The integration over $p_{z}$ and $p_{\perp}$ can easily be performed and yields
\begin{eqnarray}\label{appB10}
\int\frac{dp_{z}p_{\perp}dp_{\perp}}{\left(2\pi\right)^{2}}e^{-\left(p_{z}^{2} +p_{\perp}^{2}\right)t}=\frac{1}{(2\pi)^{3}}\left(\frac{\pi}{t}\right)^{3/2}.
\end{eqnarray}
We first substitute \eqref{appB10} into \eqref{appB8} and then perform the integration over $t$
by making use of
\begin{eqnarray}\label{appB11}
\hspace{-0.5cm}\int_{0}^{\infty }dt\,t^{x-1}e^{-w^{2}t}=\Gamma \left(x\right)\left(w^{2}\right)^{-x},
\end{eqnarray}
for $\text{Re}[w^{2}]>0$, and Re$[x]>$0.
Using, at this stage, the Legendre formula for the $\Gamma(z)$ function in \eqref{appB8}
\begin{eqnarray}\label{appB12}
\Gamma(z)=\frac{2^{z}}{2\sqrt{\pi}}\Gamma\left(\frac{z}{2}\right)\Gamma \left( \frac{z+1}{2} \right),
\end{eqnarray}
and plugging \eqref{appB11} and \eqref{appB12}, with $x=z/2$ and $w=m$, into \eqref{appB8}, we arrive at
\begin{eqnarray}\label{appB13}
\mathcal{I}_{1}&=&-\frac{m^{3}}{16\pi^{2}}\frac{1}{2\pi i}\int_{c-i\infty }^{c+i\infty }dz\,\Gamma \left( \frac{z+1}{2} \right)\Gamma \left( \frac{z-3}{2} \right)\nonumber\\
&&\times\zeta \left( 1+z \right)\left( \frac{m\beta}{2} \right)^{-z},
\end{eqnarray}
that leads to
\begin{eqnarray}\label{appB14}
\mathcal{I}_{1}&=&-\frac{\pi^{2} }{90}T^{3}+\frac{m^{2}T}{24}-\frac{m^{3}}{12\pi}+\frac{m^{4}}{32\pi^{2}T}\nonumber\\
&&\times \left( \ln \left( \frac{4\pi T }{m } \right)-\gamma_{\text{E}}+\frac{3}{4} \right)+\cdots,
\end{eqnarray}
upon using Cauchy's theorem and summing over the residues of $\Gamma$s and $\zeta$-function.
\par
Let us now consider the $\Omega$ dependent part of $V_{\text{eff}}^{(0)T}$, $\mathcal{I}_{2}$ from \eqref{appB3}. Using \eqref{appB4}, it is first given by
\begin{eqnarray}\label{appB15}
\mathcal{I}_{2}=-2\sum_{\ell=1}^{+\infty}\int\frac{dp_{z}p_{\perp}dp_{\perp}}{(2\pi)^{2}}\left(\frac{e^{-\beta\omega k}}{k}\cosh\left(k\ell\Omega \beta\right)\right). \nonumber\\
\end{eqnarray}
Expanding $\cosh\left(k\ell\Omega\beta\right)$ in \eqref{appB15}, $\mathcal{I}_{2}$ is given by
\begin{eqnarray}\label{appB16}
\mathcal{I}_{2}=-2\sum\limits_{\ell=1}^{+\infty}\sum_{j=1}^{+\infty}\frac{1}{(2j)! }\left(\beta\ell\Omega\right)^{2j}\mathcal{Q}_{j},
\end{eqnarray}
with
\begin{eqnarray}\label{appB17}
\mathcal{Q}_{j}\equiv\sum_{k=1}^{+\infty}\int\frac{dp_{z}p_{\perp}dp_{\perp}}{(2\pi)^{2}}e^{-\beta\omega k}k^{2j-1}.
\end{eqnarray}
Following the same steps leading to $\mathcal{I}_{1}$, we arrive first at
\begin{eqnarray}\label{appB18}
\mathcal{Q}_{j}&=&\frac{m^{3}}{16\pi^2}\frac{1}{2\pi i}\int_{c-i\infty}^{c+i\infty}dz~\Gamma\left(\frac{z+1}{2}\right)\nonumber\\
&&\times\Gamma\left(\frac{z-3}{2}\right)\zeta\left(1+z-2j\right)\left(\frac{m\beta}{2}\right)^{-z},
\end{eqnarray}
and then after performing the integration over $z$ by using the Cauchy's theorem, we obtain
\begin{eqnarray}\label{appB19}
\mathcal{Q}_{1}=\frac{m^{4}}{16\pi^{2}T}\zeta^{\prime}(-2)+\frac{m^{2}}{8\pi^{2}T}-\frac{m}{4\pi T^{2}}+\frac{1}{6T^{3}}+\cdots, \nonumber\\
\end{eqnarray}
and
\begin{eqnarray}\label{appB20}
\mathcal{Q}_{2}=\frac{m^{4}}{16\pi^{2}T}\zeta^{\prime}(-4)-\frac{1}{2\pi^{2}T^{3}}+\cdots.
\end{eqnarray}
Substituting these results into $\mathcal{I}_{2}$, we arrive at
\begin{eqnarray}\label{appB21}
\mathcal{I}_{2}=-\sum_{\ell=1}^{+\infty}\left(\frac{(3m^{2}-\left(\ell\Omega\right)^{2})}{24\pi^{2}T}-\frac{m}{4\pi}+\frac{1}{6T}\right)\left(\ell\Omega \right)^{2}+\cdots.\nonumber\\
\end{eqnarray}
Adding this expression with $\mathcal{I}_{1}$ from \eqref{appB14}, according to \eqref{appB2}, the HTE of $V_{\text{eff}}^{(0)T}$ is given by \eqref{A15}.
\section{High temperature expansion of $\bs{\Pi_{1}}$}\label{appC}
\setcounter{equation}{0}
As it is described in Sec. \ref{sec3B}, the one-loop self-energy function of the CKG field is given by \eqref{A28}, with $\mathcal{J}_{i}, i=1,2$ from \eqref{A29}. In this appendix, we use the method introduced in Appendix \ref{appB} and derive the HTE of $\mathcal{J}_{i}, i=1,2$.
\par
Let us first consider $\mathcal{J}_{1}$ and replace the Bose-Einstein distribution function $n_{b}(\omega)$ with
\begin{eqnarray}\label{appC1}
n_{b}(\omega)=T\frac{d}{d\alpha}\ln\left(1-e^{-\beta(\omega+\alpha)}\right)\bigg|_{\alpha=0}.
\end{eqnarray}
We thus arrive at
\begin{eqnarray}\label{appC2}
\mathcal{J}_{1}=T\frac{d}{d\alpha}\int\frac{dp_{z}p_{\perp}dp_{\perp }}{\left( 2\pi  \right)^{2}}\frac{1}{\omega }\ln \left( 1-e^{-\beta \left( \omega +\alpha  \right)} \right) \bigg|_{\alpha=0}. \nonumber\\
\end{eqnarray}
Using \eqref{appB4} to expand the logarithm in \eqref{appC2}, substituting the Mellin transformation of the exponential function \eqref{appB6} into the resulting expression, and using
\begin{eqnarray}\label{appC3}
\left(\omega^{2}\right)^{-(z+1)/2}=\frac{1}{\Gamma(z/2)}\int_{0}^{\infty}dt t^{\frac{(z+1)}{2}-1}e^{-\omega^{2}t},
\end{eqnarray}
with $\omega^{2}=p_{z}^{2}+p_{\perp}^{2}+m^{2}$, we arrive at
\begin{eqnarray}\label{appC4}
\lefteqn{\mathcal{J}_{1}=}\nonumber\\
&&-\frac{T}{2\pi i}\frac{d}{d\alpha}\int_{c-i\infty}^{c+i\infty}dz \mbox{Li}_{z+1}\left(e^{-\alpha\beta}\right)\frac{\Gamma(z)}{\Gamma((z+1)/2)}\beta^{-z}\nonumber\\
&&\times \int_{0}^{\infty} dt t^{\frac{z+1}{2}-1}e^{-m^{2}t}\int \frac{dp_{z}p_{\perp}dp_{\perp}}{(2\pi)^{2}}e^{-\left(p_{z}^{2}+p_{\perp}^{2}\right)t}\bigg|_{\alpha=0},\nonumber\\
\end{eqnarray}
where the polylogarithm $\mbox{Li}_{z+1}\left(e^{-\alpha\beta}\right)$ arises from
\begin{eqnarray}\label{appC5}
\sum_{k=1}^{+\infty}e^{-\beta k\alpha}k^{-(1+z)}=\mbox{Li}_{z+1}\left(e^{-\alpha\beta}\right).
\end{eqnarray}
Performing the integration over $p_{z}$ and $p_{\perp}$ by using \eqref{appB10}, plugging the resulting expression into \eqref{appC4}, and performing the integration over $t$ by using \eqref{appB11},
\begin{eqnarray}\label{appC6}
\mathcal{J}_{1}&=&\frac{m^{2}}{16\pi^{2}}\frac{1}{2\pi i}\int_{c-i\infty}^{c+i\infty}dz\Gamma\left(\frac{z}{2}\right)\Gamma\left(\frac{z-2}{2}\right)\mbox{Li}_{z}(1)\nonumber\\
&&\times\left(\frac{m\beta}{2}\right)^{-z}.
\end{eqnarray}
To arrive at \eqref{appC6}, we also used \eqref{appB12} and
$$\frac{d}{d\alpha}\mbox{Li}_{z+1}\left(e^{-\alpha\beta}\right)\bigg|_{\alpha=0}=-\beta\mbox{Li}_{z}(1).$$
Using Cauchy's theorem and summing over residues of $\Gamma$ and polylogarithm functions, $\mathcal{J}_{1}$
is given by
\begin{eqnarray}\label{appC7}
\mathcal{J}_{1}&=&\frac{T^{2}}{12}-\frac{mT}{4\pi}+\frac{m^{2}}{8\pi^{2}}\left(\ln\left(\frac{4\pi T}{m}\right)-\gamma_{E}+\frac{1}{2}\right)+\cdots.\nonumber\\
\end{eqnarray}
Let us now consider $\mathcal{J}_{2}$ from \eqref{A29}. To evaluate it, we use
\begin{eqnarray}\label{appC8}
n_{b}\left(\omega\pm\ell\Omega\right)=\pm T\frac{\partial}{\partial (\ell\Omega)}\ln\left(1-e^{-\beta\left(\omega\pm\ell\Omega\right)}\right),
\end{eqnarray}
and Taylor expand the logarithms according to \eqref{appB4}. We arrive at
\begin{eqnarray}\label{appC9}
\mathcal{J}_{2}&=&2T\sum_{\ell=1}^{+\infty}\sum_{k=1}^{+\infty}\frac{\partial}{\partial(\ell\Omega)}\int\frac{dp_{z}p_{\perp}dp_{\perp}}{(2\pi)^{2}}\frac{1}{\omega}\nonumber\\
&&\times \left(\frac{e^{-\beta\omega k}}{k}\sinh\left(k\ell\Omega \beta\right)\right).
\end{eqnarray}
Expanding $\sinh(k\ell\Omega\beta)$ in the orders of $\ell\Omega$, we arrive first at
\begin{eqnarray}\label{appC10}
\mathcal{J}_{2}=2\sum_{\ell=1}^{+\infty}\sum_{j=0}^{+\infty}\frac{1}{(2j)!}\mathcal{F}_{j},
\end{eqnarray}
with
\begin{eqnarray}\label{appC11}
\mathcal{F}_{j}\equiv\sum_{k=1}^{+\infty}\int\frac{dp_{z}p_{\perp}dp_{\perp}}{(2\pi)^{2}}
\frac{e^{-\beta\omega k}k^{2j}}{\omega}.
\end{eqnarray}
Following, at this stage, the same steps leading to $\mathcal{I}_{1}$ from Appendix \ref{appB}, we arrive first at
\begin{eqnarray}\label{appC12}
\mathcal{F}_{j}&=&\frac{m^{2}}{16\pi^{2}}\frac{1}{2\pi i}\int_{c-i\infty}^{c+i\infty}dz~\Gamma\left(\frac{z-2}{2}\right)\Gamma\left(\frac{z}{2}\right)\zeta\left(z-2j\right)\nonumber\\
&&\times \left(\frac{m\beta}{2}\right)^{-z},
\end{eqnarray}
and then after performing the integration over $z$ by using the Cauchy's theorem, we obtain
\begin{eqnarray}\label{appC13}
\mathcal{J}_{2}&=&\sum\limits_{\ell=1}^{+\infty}\bigg[
\frac{T^{2}}{6}-\frac{(2m^{2}-(\ell\Omega)^{2})T}{4\pi m}-\frac{(\ell\Omega)^{2}}{4\pi^{2}}\nonumber\\
&&+\frac{m^{2}}{8\pi^{2}}\left(\ln\left(\frac{4\pi T}{m}\right)-\gamma_{E}+\frac{1}{2}\right)
\bigg].
\end{eqnarray}
Adding this expression to $\mathcal{J}_{1}$ from \eqref{appC7}, we arrive, according to \eqref{A28} at $\Pi_{1}^{\text{mat}}$ from \eqref{A30}.
\end{appendix}


\end{document}